\documentclass[aps,prl,superscriptaddress,twocolumn]{revtex4-1}

\usepackage{graphicx}
\usepackage{dcolumn}
\usepackage{amssymb}
\usepackage{amsmath}
\usepackage{bbold}
\usepackage{bm}
\usepackage{color}
\usepackage{physics}
\usepackage{verbatim}

\usepackage{soul}

\usepackage[colorlinks,urlcolor=blue,citecolor=blue,linkcolor=blue]{hyperref}
\usepackage{comment}
\usepackage{cleveref}
\usepackage{mathtools}
\usepackage{mathrsfs}

\newcommand{\sch}{Schr{\"o}dinger }

\renewcommand{\k}{\mathbf{k}}

\newcommand{\0}{\mathbf{0}}

\newcommand{\q}{\mathbf{q}}

\newcommand{\kprime}{\mathbf{k}^\prime}

\makeatletter
\newcommand{\subalign}[1]{%
  \vcenter{%
    \Let@ \restore@math@cr \default@tag
    \baselineskip\fontdimen10 \scriptfont\tw@
    \advance\baselineskip\fontdimen12 \scriptfont\tw@
    \lineskip\thr@@\fontdimen8 \scriptfont\thr@@
    \lineskiplimit\lineskip
    \ialign{\hfil$\m@th\scriptstyle##$&$\m@th\scriptstyle{}##$\hfil\crcr
      #1\crcr
    }%
  }%
}
\makeatother

\begin{document}

\title{Light-enhanced dipolar interactions between exciton polaritons}

\author{Yasufumi Nakano}
\affiliation{School of Physics and Astronomy, Monash University, Victoria 3800, Australia}

\author{Olivier Bleu}
\affiliation{School of Physics and Astronomy, Monash University, Victoria 3800, Australia}
\affiliation{Institut f{\"u}r Theoretische Physik, Universit{\"a}t Heidelberg, 69120 Heidelberg, Germany}

\author{Brendan C. Mulkerin}
\affiliation{School of Physics and Astronomy, Monash University, Victoria 3800, Australia}

\author{Jesper Levinsen}
\affiliation{School of Physics and Astronomy, Monash University, Victoria 3800, Australia}

\author{Meera M. Parish}
\affiliation{School of Physics and Astronomy, Monash University, Victoria 3800, Australia}

\date{\today}

\begin{abstract}
We consider the scenario of excitons in a semiconductor bilayer that are strongly coupled to cavity photons, leading to the formation of dipolar exciton polaritons (dipolaritons). Using a realistic pseudopotential for the dipolar interactions, we exactly determine the scattering between dipolaritons, accounting for the hybridization between interlayer and intralayer excitons. Similar to conventional non-dipolar polaritons, we find that the light-matter coupling enhances the interactions between dipolaritons by forcing excitons to scatter at energies that would otherwise be forbidden in ordinary exciton-exciton collisions. However, we show that this light enhancement is larger for long-range dipolar interactions than for short-range intralayer interactions, and is sensitive to the (non-uniform) dielectric environment of the bilayer. Crucially, we find that the largest dipolariton interactions are achieved for transition metal dichalcogenide bilayers in vacuum. Our results thus reveal the optimal dipolariton setup for realizing strong photon correlations.
\end{abstract}

\maketitle

Exciton polaritons are hybrid light-matter quasiparticles that arise when semiconductor excitons (bound electron-hole pairs) are strongly coupled to a photon mode in an optical microcavity~\cite{Kavokin2017microcavities,Carusotto2013}. Owing to their bosonic nature and small effective mass, a variety of collective coherent phenomena, including Bose-Einstein condensation~\cite{Kasprzak2006,Balili2007,Deng2010} and superfluidity~\cite{Amo2009,Lagoudakis2008,Sanvitto2010,Lerario2017}, has been observed. Moreover, the ability to control polaritons using potential landscapes and/or their polarization has enabled the observation of topological phenomena~\cite{StJean2017,Klembt2018,Gianfrate2020,Pieczarka2021,Solnyshkov2021}, as well as the potential realization of optoelectronic devices~\cite{Wertz2012,Gao2012,Ballarini2013,Li2024,Sanvitto2016}. These effects are, however, mainly semiclassical and the realization of correlated quantum effects such as the polariton blockade~\cite{Verger2006} has proven difficult due to the relatively weak polariton-polariton interactions in typical experiments~\cite{MunozMatutano2019,Delteil2019}. Achieving strong correlations between polaritons would open new perspectives for quantum photonic applications in these scalable semiconductor systems~\cite{Gerace2009,Gerace2019,Liew2023}.

\begin{figure}[tp]
    \centering
    \begin{minipage}{0.48\textwidth}
    \centering
    \includegraphics[width=\textwidth]{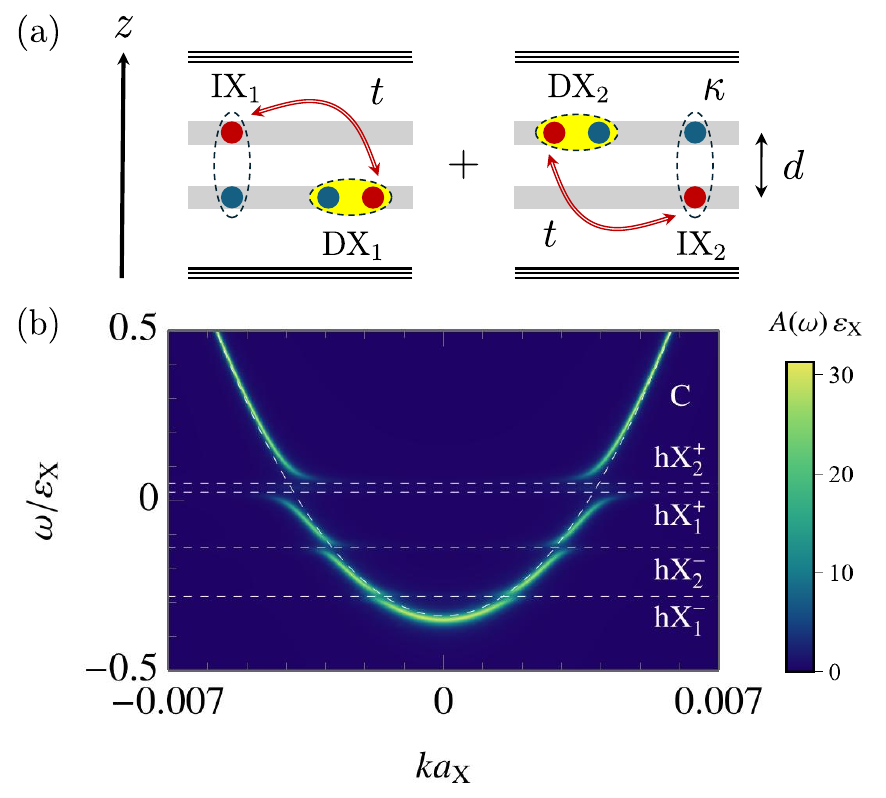}
    \end{minipage} 
    \caption{(a) Schematic of the dipolariton in a bilayer. Direct excitons DX$_1$ and DX$_2$ are coupled to the cavity photon (yellow region), while each DX hybridizes with the corresponding indirect exciton IX via tunneling $t$ of the hole (red). The IX dipole moment depends on the interlayer separation $d$ and the dielectric constant $\kappa$ of the surrounding environment. For co-polarized photons, DX$_1$ is in the $K'$ valley and DX$_2$ is in the $K$ valley (see text). (b) Photon spectral function, featuring the five polariton branches. We also show the photon (C) and hybrid excitons (hX$^+$ and hX$^-$) in the absence of light-matter coupling (dashed white). We use parameters inspired by MoS$_2$ homobilayer experiments~\cite{Leisgang2020,Lorchat2021,Datta2022,Louca2023}: $\Omega/\varepsilon_\text{X} = 0.085$, $t/\varepsilon_\text{X} = 0.17$, $\delta_\text{C}/\varepsilon_\text{X} = -0.34$, $\delta_\text{IX}/\varepsilon_\text{X} = -0.17$, $\Delta/\varepsilon_\text{X} = 0.085$, and $\Gamma/\varepsilon_\text{X} = 0.01$, where the reference scales $a_\text{X}$ and $\varepsilon_\text{X}$ correspond to $\kappa=1$ (see text).}
    \label{fig:schematic}
\end{figure}

A promising route towards enhancing the polariton-polariton interactions is to exploit excitons with long-range dipolar interactions---most notably, spatially indirect interlayer excitons~\cite{Rivera2015,Arora2017,Calman2018,Horng2018,Niehues2019,Sun2024}. However, the challenge is to maintain a strong coupling to light while supporting dipole-dipole interactions, since a large dipole moment necessarily requires a sizeable electron-hole separation, which renders the exciton state optically dark. Hybrid interlayer excitons~\cite{Deilmann2018,Gerber2019,Leisgang2020,Lorchat2021} have thus emerged as an appealing candidate for dipolar polaritons (dipolaritons), since they inherit a strong oscillator strength from the intralayer excitons, while acquiring a permanent dipole moment from the interlayer exciton [Fig.~\ref{fig:schematic}(a)]. Experiments have already observed the formation of dipolaritons in bilayers~\cite{Cristofolini2012,Togan2018,Datta2022,Louca2023}; however, there is currently a lack of theory that can describe the significant enhancement of interactions reported in both GaAs-based quantum wells~\cite{Togan2018} and transition metal dichalcogenides (TMDs)~\cite{Datta2022,Louca2023}.

In this Letter, we present the first non-perturbative theory of dipolariton interactions for the experimentally important case of hybrid excitons in a two-dimensional (2D) bilayer, accounting for the influence of the non-uniform dielectric environment (Fig.~\ref{fig:schematic}). We determine all the polariton-polariton scattering processes, starting from pseudopotentials for the underlying excitonic interactions that are informed by fully microscopic approaches~\cite{Ciuti1998,Tassone1999,Byrnes2010}. Crucially, we find that the dipolariton interactions can be substantially enhanced compared to those of dipolar excitons, since the coupling to light forces excitons to scatter at energies that would otherwise be inaccessible. Notably, such energy-dependent scattering is absent in standard perturbative theories of the interactions for hybrid interlayer excitons~\cite{Nalitov2014a,Maslova2024} and dipolaritons~\cite{Byrnes2014,Nalitov2014b}, which were conducted at the level of the first- or second-order Born approximation. Our work also goes beyond recent numerical calculations for two dipolaritons~\cite{Christensen2024}, which were restricted to the idealized scenario of one dimension and pointlike dipolar excitons. 

While a similar light-enhanced interaction has been predicted for conventional polaritons with underlying short-range interactions~\cite{Bleu2020}, we show that the effect is even larger for long-range interactions and controlled by the surrounding dielectric environment through the dipole moment. Furthermore, we find that the dark exciton configuration in the bilayer leads to a further enhancement of dipolariton interactions compared to those of conventional polaritons. 

\paragraph{Model.---}
We consider the scenario of dipolaritons in an optical microcavity shown in Fig.~\ref{fig:schematic}(a). Here, we have a superposition of a cavity photon, two direct excitons (DXs) formed inside two spatially separated layers, and two indirect excitons (IXs) formed across the two layers with static out-of-plane electric dipole moments pointing in opposite directions.

While our model is applicable to a broad range of materials, for concreteness, we focus on a naturally stacked 2H (AA$'$) MoS$_2$ homobilayer as in many experiments~\cite{Leisgang2020,Lorchat2021,Datta2022,Louca2023}, and assume co-circularly polarized excitons. Within this configuration, we label the relevant exciton modes by a single index $\eta \in \{1,2\}$ that fixes the layer/valley for DXs and the valley/dipole orientation for IXs. Specifically, we take $\eta=1$ to denote the DX in the bottom layer and $K'$ valley, and $\eta=2$ the DX in the top layer and $K$ valley. The associated IX mode carries the same valley label $\eta$ and has an out-of-plane dipole moment along $+z$ for $\eta=1$ and $-z$ for $\eta=2$, such that the applied electric field produces opposite Stark shifts for the two branches. We define $+z$ to point from the bottom layer to the top layer, as indicated in Fig.~\ref{fig:schematic}(a).

Setting $\hbar$ and the area to 1, the non-interacting Hamiltonian describing this system takes the form
\begin{align}
    &\hat{H}_{0} = \sum_{\k} \Big[ \epsilon^{\text C}_{\k}\hat{c}^{\dagger}_{\k}\hat{c}_{\k} + \sum^{2}_{\eta=1}\Big( \epsilon^{\text{DX}}_{\k}\hat{x}^{\dagger}_{\k,\eta}\hat{x}_{\k,\eta} + \epsilon^{\text{IX}}_{\k,\eta}\hat{y}^{\dagger}_{\k,\eta}\hat{y}_{\k,\eta} \notag \\
    &\quad + \frac{\Omega}{2}\big( \hat{x}^{\dagger}_{\k,\eta}\hat{c}_{\k} + \text{h.c.}\big) + \frac{t}{2}\big( \hat{x}^{\dagger}_{\k,\eta}\hat{y}_{\k,\eta} + \text{h.c.} \big) \Big) \Big].
    \label{eq:Hamiltonian}
\end{align}
Here, $\hat{c}_{\k}$ denotes the annihilation operator for a cavity photon with dispersion $\epsilon^\text{C}_{\k} = k^2/(2m_\text{C}) + \delta_\text{C}$, where $\delta_\text{C}$ is the photon-DX detuning. $\hat{x}_{\k,\eta}$ denotes the annihilation operator for a DX with dispersion $\epsilon^{\text{DX}}_{\k} = k^2/(2m_\text{DX})$. $\hat{y}_{\k,\eta}$ denotes the annihilation operator for an IX with dispersion $\epsilon^{\text{IX}}_{\k,\eta}=k^2/(2m_\text{IX})+\delta_\text{IX}\mp\Delta$, where $\delta_\text{IX}$ is the bare IX-DX detuning and $\Delta$ is the Stark shift arising from an applied out-of-plane electric field~\cite{Leisgang2020,Lorchat2021}. The sign in front of $\Delta$ is taken to be $-$ for $\eta=1$ and $+$ for $\eta=2$.

The DXs are strongly coupled to the cavity photon via the Rabi splitting $\Omega$, while the IXs are coupled to the corresponding DX through the delocalization of the hole via the tunneling rate $t$. We assume that the Rabi splitting and the tunneling rate are not affected by the dielectric environment, since the DX and IX wave functions, which are responsible for these parameters, are relatively insensitive to the dielectric environment~\cite{Maslova2024}. We further assume that the DX and IX have equal masses $m_\text{DX} = m_\text{IX} = m_\text{X}$, and we take the cavity photon mass $m_\text{C} = 10^{-5}m_\text{X}$.

In the absence of light-matter coupling, the hole tunneling leads to DX$_{\eta}$-IX$_{\eta}$ hybridization for $\eta=1,2$. The upper (hX$^+$) and lower (hX$^-$) hybrid exciton dispersions are given by
\begin{equation}
    E^{\pm}_{\k,\eta} = \frac{1}{2}\left( \epsilon^{\text{IX}}_{\k,\eta} + \epsilon^{\text{DX}}_{\k} \pm \sqrt{(\epsilon^{\text{IX}}_{\k,\eta}-\epsilon^{\text{DX}}_{\k})^{2} + t^{2}} \right).
    \label{eq:hybrid exciton dispersions}
\end{equation}
Incorporating the light-matter coupling then generates exciton polaritons, where the polariton dispersions are obtained by diagonalizing $\hat H_0$ in Eq.~\eqref{eq:Hamiltonian} as
\begin{equation}
    \hat{H}_{0} = \sum_{\k}\sum^{5}_{n=1} E^{(n)}_{\k}\hat{P}^{\dagger}_{\k,n}\hat{P}_{\k,n},
    \label{eq:eigenvalues}
\end{equation}
via the linear transformation
\begin{equation}
    \hat{P}_{\k,n} = C^{(n)}_{\k}\hat{c}_{\k} + X^{(n)}_{\k,1}\hat{x}_{\k,1}  + X^{(n)}_{\k,2}\hat{x}_{\k,2} + Y^{(n)}_{\k,1}\hat{y}_{\k,1} + Y^{(n)}_{\k,2}\hat{y}_{\k,2}.
    \label{eq:linear transformation}
\end{equation}
Here, $\hat{P}_{\k,n}$ corresponds to the $n$th polariton mode with $n \in \{1,2,\cdots,5\}$, whose energy satisfies $E^{(n_1)}_{\k} < E^{(n_2)}_{\k}$ for $n_{1}<n_{2}$. The transformation (Hopfield) coefficients for the photon, DX$_1$, DX$_2$, IX$_1$, and IX$_2$ are denoted $C^{(n)}_{\k}$, $X^{(n)}_{\k,1}$, $X^{(n)}_{\k,2}$, $Y^{(n)}_{\k,1}$, and $Y^{(n)}_{\k,2}$, respectively, which can be chosen to be real without loss of generality. Their squares correspond to the mode fractions for a given polariton and satisfy $(C^{(n)}_{\k})^{2} + (X^{(n)}_{\k,1})^{2} + (X^{(n)}_{\k,2})^{2} + (Y^{(n)}_{\k,1})^{2} + (Y^{(n)}_{\k,2})^{2} = 1$. 

Figure~\ref{fig:schematic}(b) shows a typical spectrum, as observed in experiment, which is obtained from the photon spectral function~\cite{Cwik2016}
\begin{equation}
    A(\omega) = -\frac{1}{\pi}\Im\left[ \bra{0}\hat{c}_{\k}\frac{1}{\omega-\hat{H}_{0}+i\Gamma}\hat{c}^{\dagger}_{\k}\ket{0} \right],
    \label{eq:spectral function}
\end{equation}
where $\Gamma$ corresponds to the cavity photon linewidth and $\ket{0}$ is the vacuum state of the microcavity. We clearly observe four polariton modes (P$_{1}$, P$_{2}$, P$_{3}$, P$_{5}$), while the remaining P$_{4}$ mode has only a small photon fraction $(C^{(4)}_{\k})^{2}$ and is thus almost invisible.

\begin{figure}[tp]
    \begin{minipage}{0.48\textwidth}
    \centering
    \includegraphics[width=\textwidth]{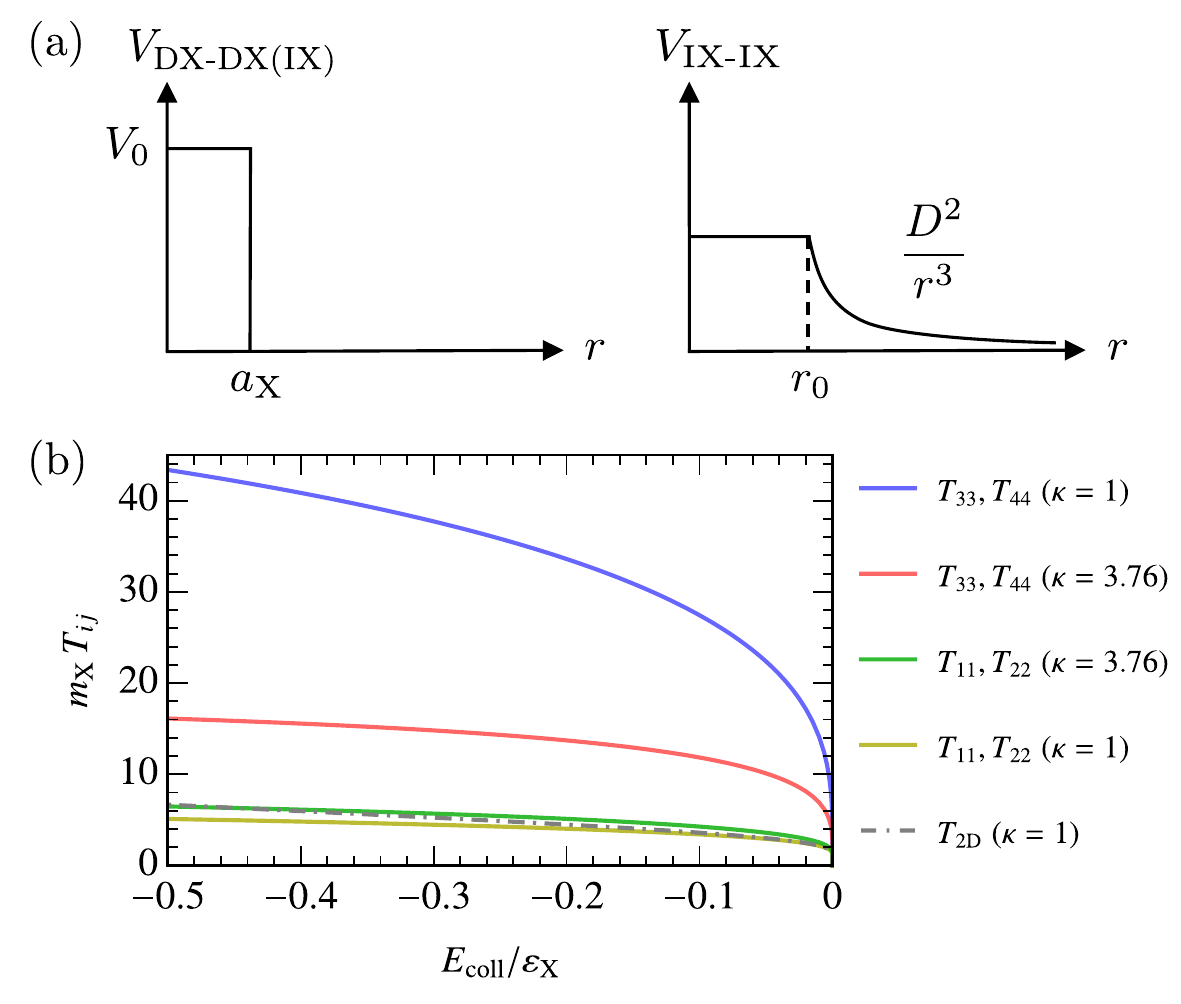}
    \end{minipage}
    \caption{(a) Schematic of the intravalley DX-DX(IX) (left) and IX-IX (right) interaction potentials, where the latter exhibits a long-range dipolar tail. (b) Exciton $T$ matrices at zero momentum and zero tunneling as a function of collision energy, i.e., the energy measured from the corresponding two-particle continuum. The blue and red lines show the IX-IX $T$ matrix $T_{33}=T_{44}$ for vacuum ($\kappa=1$) and hBN ($\kappa=3.76$), respectively. The yellow and green lines show the DX-DX $T$ matrix $T_{11}=T_{22}$ for $\kappa=1$ and $\kappa=3.76$, respectively. The gray dot-dashed line is the 2D universal low-energy expression \eqref{eq:2D T-matrix} expected for DX-DX interactions, with $a_\text{2D}/a_\text{X}=0.39$ obtained from a fit at small collision energy. In panel (b), we use $\rho_{0}/a_{0} = 40$ and $d/a_{0} = 6$, with the reference scales $a_\text{X}$, $\rho_{0}$, $a_{0}$, and $\varepsilon_\text{X}$ fixed by taking $\kappa=1$.}
    \label{fig:Tx}
\end{figure}

\paragraph{Exciton interaction potentials.---}
The potential operator describing the interactions between excitons takes the form
\begin{align}
    \hat{V} &= \frac{1}{2}\sum_{\k\kprime\q}\sum^{2}_{\eta=1} V_{\text{DX-DX}}(\q) \,\hat{x}^{\dagger}_{\k+\q,\eta}\hat{x}^{\dagger}_{\kprime-\q,\eta}\hat{x}_{\kprime,\eta}\hat{x}_{\k,\eta} \notag \\
    &+ \frac{1}{2}\sum_{\k\kprime\q}\sum^{2}_{\eta=1} V_{\text{IX-IX}}(\q) \,\hat{y}^{\dagger}_{\k+\q,\eta}\hat{y}^{\dagger}_{\kprime-\q,\eta}\hat{y}_{\kprime,\eta}\hat{y}_{\k,\eta} \notag \\
    &+ \sum_{\k\kprime\q}\sum^{2}_{\eta=1} V_{\text{DX-IX}}(\q) \,\hat{x}^{\dagger}_{\k+\q,\eta}\hat{y}^{\dagger}_{\kprime-\q,\eta}\hat{y}_{\kprime,\eta}\hat{x}_{\k,\eta} \, .
    \label{eq:potential operator}
\end{align}
To describe the interactions between excitons, we use pseudopotentials as depicted in Fig.~\ref{fig:Tx}(a), where the parameters are informed by a microscopic description that explicitly considers the underlying intra- and interlayer Rytova-Keldysh potentials~\cite{Asriyan2019,Semina2019} for the constituent electrons and holes forming the excitons. More specifically, we set the parameters such that the Born approximation for each pseudopotential matches that obtained from the microscopic description (for details, see the Supplemental Material~\cite{supmat}). We note that the Born approximation is conceptually important since it provides an upper bound on the two-body interaction strength~\cite{Li2021a}. 

In writing the potential operator in Eq.~\eqref{eq:potential operator}, we have omitted antiparallel IX-IX interactions because, within the regime of interest ($0.2 \leq \Delta/\Omega \leq 1.2$), the electric-field-induced Stark shift $\Delta$ is a sizable fraction of the Rabi splitting $\Omega$. As a result, one dipole orientation acquires a much larger IX fraction in the relevant polariton mode and therefore dominates the polariton-polariton scattering. In contrast, antiparallel IX-IX interactions are expected to become important near $\Delta \simeq 0$, where the two IX modes are nearly degenerate and both dipole orientations contribute substantially to the scattering process. Intervalley DX-DX and DX-IX interactions are likewise dropped, as their zero-momentum Born approximation vanishes~\cite{supmat}.

In order to account for the non-uniform dielectric environment of the bilayer, we introduce the dielectric screening length $\rho_{0} = 2\pi\chi_\text{2D}/\kappa$, where $\chi_\text{2D}$ denotes the 2D polarizability of the layers. The dielectric screening length and other derived quantities (such as $a_\text{X}$, $\varepsilon_\text{X}$, $r_{0}$, and $a_{0}$) are parameterized by the environmental dielectric constant $\kappa$. For brevity, we suppress their explicit $\kappa$ dependence. In the present work, we use $\kappa=1$ for vacuum and $\kappa=3.76$ for the case of hBN encapsulation~\cite{Laturia2018}.

We model intravalley DX-DX and DX-IX interactions as short-range soft-core potentials in real space
\begin{equation}
    V_\text{DX-DX(IX)}(\mathbf{r}) = V^{\text{DX-DX(IX)}}_{0}\theta\left(a_\text{X}-r\right),
\end{equation}
where $\theta(x)$ denotes the Heaviside step function. Here, the range of the potential is set by the effective size of the exciton $a_\text{X}=1/\sqrt{2\mu\varepsilon_\text{X}}$, where $\varepsilon_\text{X}$ is the DX binding energy for dielectric constant $\kappa$ and $\mu$ is the electron-hole reduced mass. For a homobilayer, we take equal electron and hole masses $\mu = m_\text{X}/4$.

For parallel IX-IX interactions, we expect the potential to feature a dipolar tail at large distances while still retaining a soft repulsive core at short distances, where the excitons overlap [see Fig.~\ref{fig:Tx}(a)]. This leads to 
\begin{equation}
    \label{eq:VIX-IX in real space}
    V_\text{IX-IX}(\mathbf{r}) = \frac{D^2}{r_{0}^{3}}\theta\left(r_{0}-r\right) + \frac{D^2}{r^{3}}\theta\left(r-r_{0}\right), 
\end{equation}
with strength $D^2 = d^{2}/(2\mu a_{0})$ due to the IX dipole moment. Here, $a_{0} = \kappa/(2\mu e^{2})$ corresponds to the effective 2D Bohr radius, where $e$ is the elementary charge. If we had taken $r_0 \to 0$, i.e., pointlike dipoles as in Ref.~\cite{Christensen2024}, then the Born approximation would diverge and no longer be defined, thus implying an unbounded interaction strength. 

The use of such effective potentials in polariton scattering is justified as long as the exciton binding energy greatly exceeds other energy scales in the problem~\cite{Bleu2020}, which is well satisfied for MoS$_2$ homobilayers. In particular, this means that we can apply effective exciton interaction potentials while neglecting any tunneling- and/or light-induced changes to the exciton wave function~\cite{Khurgin2001,Levinsen2019a}.

\paragraph{Hybrid exciton scattering.---} 
To determine the full scattering properties of hybrid excitons or dipolaritons, we must go beyond the Born approximation and sum the entire Born series. This involves considering an infinite number of scattering processes, which we sum using the Lippmann-Schwinger equation~\cite{Sakurai2020}, appropriately generalized to the case of a light-matter coupled system~\cite{Bleu2020}. Similar approaches have previously been applied to investigate non-dipolar polariton-polariton~\cite{Wouters2007,Bleu2020,Li2021c} and polariton-electron~\cite{Li2021a,Li2021b,Kumar2023} interactions. 

The central object governing the polariton interaction strength is the scattering $T$ matrix. Since interactions only occur between excitons, it is sufficient to consider only its exciton matrix elements~\cite{supmat}
\begin{align}
    &T_{ij}(k',k;E) = V_{ij}(k',k)\delta_{ij} \notag \\ 
    &\quad + \sum^{6}_{n=1}\int^{\infty}_{0}\frac{q\,dq}{2\pi} V_{ii}(k',q)G_{in}(q,E)T_{nj}(q,k;E),
    \label{eq:scattering integral equation}
\end{align}
where $i,j \in \{1,\cdots,6\}$ and we have projected onto the $s$-wave channel since we consider scattering at the low momenta relevant to polaritons. Here, $i$ ($j$) labels the outgoing (incoming) interacting two-particle state at zero center-of-mass momentum with relative momentum $\k'$ ($\k$). These states are defined as $\ket{1,\k} = \hat{x}^{\dagger}_{\k,1}\hat{x}^{\dagger}_{-\k,1}\ket{0}$, $\ket{2,\k} = \hat{x}^{\dagger}_{\k,2}\hat{x}^{\dagger}_{-\k,2}\ket{0}$, $\ket{3,\k} = \hat{y}^{\dagger}_{\k,1}\hat{y}^{\dagger}_{-\k,1}\ket{0}$, $\ket{4,\k} = \hat{y}^{\dagger}_{\k,2}\hat{y}^{\dagger}_{-\k,2}\ket{0}$, $\ket{5,\k} = \frac{1}{\sqrt{2}} ( \hat{x}^{\dagger}_{\k,1}\hat{y}^{\dagger}_{-\k,1} + \hat{y}^{\dagger}_{\k,1}\hat{x}^{\dagger}_{-\k,1} ) \ket{0}$, and $\ket{6,\k} = \frac{1}{\sqrt{2}} ( \hat{x}^{\dagger}_{\k,2}\hat{y}^{\dagger}_{-\k,2} + \hat{y}^{\dagger}_{\k,2}\hat{x}^{\dagger}_{-\k,2} ) \ket{0}$. Correspondingly, the matrix elements $V_{11(22)}=V_\text{DX-DX}$, $V_{33(44)}=V_\text{IX-IX}$, and $V_{55(66)}=V_\text{DX-IX}$, where we define $V(k',k)$ as the $s$-wave projection of the momentum-space potential: $V(k',k) = \int^{2\pi}_{0}\frac{d\theta_{\k'\k}}{2\pi}V(\k'-\k)$ with $\theta_{\k'\k}$ denoting the angle between $\k'$ and $\k$. The two-particle Green's function is defined as
\begin{align}
    G_{in}(\q,E) = \bra{i,\q}\frac{1}{(E+i0)\mathbb{1}-\hat{H}_{0}}\ket{n,\q}.
\end{align}

A key feature of polariton scattering is that the light-matter coupling effectively allows excitons to interact at energies that would normally be inaccessible. To see why this is an advantage, we show in Fig.~\ref{fig:Tx}(b) the DX-DX and IX-IX $T$ matrices at zero momentum and negative collision energy $E_\text{coll}$ in the absence of light-matter coupling~\cite{supmat}. Here, to focus on the effect of dipolar interactions, we have excluded the DX-IX $T$ matrix, which yields a much smaller contribution to dipolariton scattering than the terms we have kept~\cite{supmat}. We also take $t=0$ to investigate the two types of exciton interactions separately. We clearly see that the interaction strength increases dramatically as we increase the negative collision energy for both DX and IX scattering. Moreover, we observe that the dipolar interactions are consistently larger than those of DXs, even for the interlayer separation achievable in current MoS$_2$ homobilayer experiments~\cite{Leisgang2020,Lorchat2021,Datta2022,Louca2023}. These results indicate that scattering can be enhanced by light-matter coupling, forcing excitons to scatter in the ``off-shell'' regime~\cite{Morgan2002} where the $T$ matrix is enhanced. Finally, in agreement with previous results within the Born approximation~\cite{Maslova2024}, this enhancement will be particularly pronounced when the polaritons feature a large dipolar component in vacuum.

In Fig.~\ref{fig:Tx}(b) we also show the universal low-energy $T$ matrix for two excitons in a 2D geometry~\cite{Adhikari1986}
\begin{equation} \label{eq:2D T-matrix}
    T_\text{2D}(E_\text{coll}) = \frac{4\pi}{m_\text{X}}\frac{1}{\ln(-\varepsilon_{a}/(E_\text{coll}+i0))}.
\end{equation}
Here, $\varepsilon_{a} = 1/(m_\text{X}a^{2}_\text{2D})$ is the energy scale associated with the scattering length $a_\text{2D}$ between two excitons. We see that this expression captures the DX-DX scattering very well. In principle, it can also be used for dipolar interactions with a modified scattering length; however, it quickly fails when moving away from zero collision energy since the interactions are long range~\cite{Hofmann2021}.

\begin{figure}[tp]
    \centering
    \begin{minipage}{0.48\textwidth}
    \centering
    \includegraphics[width=\textwidth]{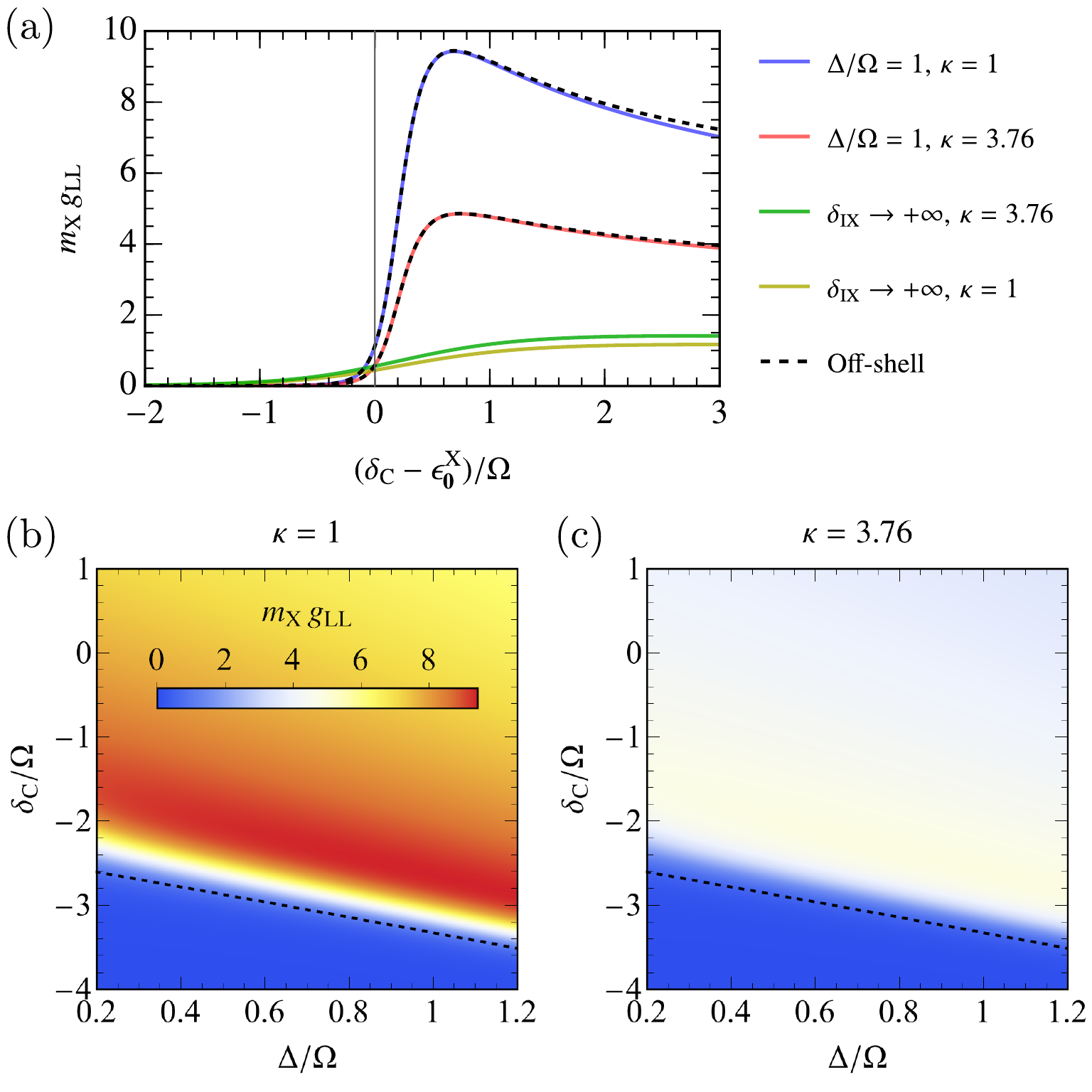}
    \end{minipage} 
    \caption{(a) LP-LP interaction constant at zero momentum as a function of photon detuning, where $\epsilon^\text{X}_\mathbf{0}=E^{-}_{\mathbf{0},1}$ for dipolaritons and $\epsilon^\text{X}_\mathbf{0}=0$ for conventional polaritons. The blue and red lines correspond to dipolaritons for vacuum ($\kappa=1$) and hBN ($\kappa=3.76$), respectively, with $\Delta/\Omega = 1$. The black dashed lines show the corresponding off-shell approximations. The limit $\delta_\text{IX}\to+\infty$ (yellow and green lines) corresponds to conventional polaritons. (b,c) Density plots of the LP-LP interaction constant for $\kappa=1$ and $\kappa=3.76$, respectively. The additional black dashed line indicates $\delta_\text{C} = E^{-}_{\mathbf{0},1}$. In all panels, we use $\rho_{0}/a_{0} = 40$, $d/a_{0} = 6$, $\Omega/\varepsilon_\text{X} = 0.085$,  $t/\varepsilon_\text{X} = 0.17$, and we use $\delta_\text{IX}/\varepsilon_\text{X} = -0.17$ unless otherwise specified. The reference scales $\rho_{0}$, $a_{0}$, and $\varepsilon_\text{X}$ correspond to $\kappa=1$.}
    \label{fig:gLL}
\end{figure}

\paragraph{Dipolariton interaction constant.---}
As we now show, the strong energy dependence of exciton scattering can lead to a large enhancement of dipolariton scattering. To be concrete, we consider the scattering between two lower polaritons (LPs), i.e., the P$_{1}$ modes, in the limit $k',k\to0$, which is the focus of most studies; however, our methodology translates directly to all the other branches and to finite momentum. To obtain the LP interaction constant, we use the $T$ matrix in Eq.~\eqref{eq:scattering integral equation}, projected onto the excitonic interaction channels
\begin{equation}
    g_{\text{LL}} = \sum^{6}_{i,j=1} \bra{P_{1},P_{1},0}\ket{i,0}\bra{j,0}\ket{P_{1},P_{1},0} T_{ij}(0,0;2E^{(1)}_{0}),
    \label{eq:LP interaction constant}
\end{equation}
where $\ket{P_{1},P_{1},\k} = \hat{P}_{1,\k}^{\dagger}\hat{P}_{1,-\k}^{\dagger}\ket{0}$. Here, the Hopfield coefficients and $T$-matrix elements are evaluated at zero momentum, but crucially, the energy is $E=2E^{(1)}_\mathbf{0}$ which is below that of any two excitons [Fig.~\ref{fig:schematic}(b)] and hence corresponds to a negative collision energy. The interaction constant strongly depends on the DX and IX fractions of the LP mode, which can both be tuned via the photon detuning and an applied out-of-plane electric field~\cite{supmat}.

Figure~\ref{fig:gLL}(a) shows our calculated polariton interaction constant as a function of photon detuning. We clearly see that placing the bilayer in vacuum can greatly enhance the interactions. In particular, when comparing to the case of a conventional bilayer without a dipolar component (corresponding to $\delta_\text{IX}\to+\infty$), we find that this enhancement can increase by nearly a factor of 8. This significant enhancement is in part because, unlike conventional polaritons in a bilayer~\cite{Bleu2020}, the dipolariton in the presence of an applied out-of-plane electric field does not suffer from a factor of 2 reduction in interactions due to the lifted degeneracies of the two IX modes. Instead, it purely benefits from the $\sqrt{2}$ increase in the Rabi coupling in the bilayer.
 
As expected, we see that $g_{\text{LL}}$ is small when the polariton is primarily photonic (i.e., when $\delta_\text{C}<E^{-}_{\mathbf{0},1}$); however, what may be less intuitive is that the interactions decrease at large positive photon detuning with respect to the hX$^{-}_{1}$ mode. Indeed, this effect arises from the strong energy dependence of the interactions [Fig.~\ref{fig:Tx}(b)], since the light-induced collision energy becomes small in this regime. This is a strong qualitative difference from the Born approximation~\cite{Byrnes2014,Nalitov2014b}, which does not contain a similar energy dependence, and it illustrates that the largest polariton interactions occur when there is a significant IX component while still maintaining a non-negligible photon fraction.

In Figs.~\ref{fig:gLL}(b,c), we explore this effect further by showing the polariton interaction constant as a function of both photon detuning and Stark shift. This clearly illustrates that, in order to have strongly enhanced scattering, we simultaneously need $\delta_\text{C} \gtrsim E^{-}_{\mathbf{0},1}$ and a sufficiently strong electric field such that the LP has both a significant dipolar and photonic fractions. Our results are in qualitative agreement with the experiments of Ref.~\cite{Togan2018}, which found a strongly enhanced LP interactions as the IX mode approached the photon mode from above.

Finally, we provide unequivocal evidence that the main role of the light-matter coupling is to force the excitons to scatter at forbidden ``off-shell'' energies, as in Fig.~\ref{fig:Tx}(b), without modifying the excitonic interactions themselves. This implies that we can, to a good approximation, evaluate Eq.~\eqref{eq:scattering integral equation} in the absence of light-matter coupling, but with a collision energy set by the polariton energies~\cite{supmat}. As shown in Fig.~\ref{fig:gLL}(a), this off-shell approximation is in excellent agreement with the exact calculations over a large range of photon detunings. Deviations are only observed for very excitonic detunings, where the polariton becomes difficult to address optically.

\paragraph{Conclusion and outlook.---}
We have formulated a non-perturbative theory of dipolariton scattering that shows how both the dielectric environment and the light-matter coupling can be used to enhance the dipolar interactions at low momentum. Our results thus indicate that hybrid interlayer excitons in TMDs provide a viable route towards realizing strongly correlated polaritons. We have furthermore identified the light-induced shift in collision energy as being key to determining the optimal conditions for strong interactions. Notably, this feature is absent in previous theories based on the standard Born approximation.

Our approach can naturally be extended to other scenarios that are relevant to experiments, such as dipolar excitons with different spins~\cite{Schindler2008,Lee2009} and electrically induced nonlinearities in a single layer~\cite{Tsintzos2018,Rosenberg2016,Rosenberg2018,Dror2024}. In this context, a key open question is how far dipolariton interactions can be enhanced by electric-field engineering of the hybridization and dipole moment.

\acknowledgments 
We acknowledge useful discussions with Shimpei Endo, Michael Fraser, Sangeet Kumar, and Pascal Naidon. We acknowledge support from the Australian Research Council Centre of Excellence in Future Low-Energy Electronics Technologies (CE170100039). JL and MMP are also supported through Australian Research Council Discovery Project DP240100569 and Future Fellowship FT200100619, respectively.

\bibliography{condensed-matter}

\newpage

\onecolumngrid
\clearpage
\begin{center}
\textbf{\large Supplemental Material:
			Light-enhanced dipolar interactions between exciton polaritons}\\
\vspace{4mm}
{Yasufumi~Nakano$^{1}$, Olivier~Bleu$^{1,2}$, Brendan C. Mulkerin$^{1}$, Jesper~Levinsen$^{1}$, and Meera~M.~Parish$^{1}$}\\
\vspace{2mm}
{\em \small
$^1$School of Physics and Astronomy, Monash University, Victoria 3800, Australia\\
$^2$Institut f{\"u}r Theoretische Physik, Universit{\"a}t Heidelberg, 69120 Heidelberg, Germany
}\end{center}
\setcounter{equation}{0}
\renewcommand{\theequation}{S\arabic{equation}}
\renewcommand{\theHequation}{S\arabic{equation}}

\setcounter{figure}{0}
\renewcommand{\thefigure}{S\arabic{figure}}
\renewcommand{\theHfigure}{S\arabic{figure}}

\setcounter{table}{0}
\setcounter{page}{1}
\makeatletter

\section{Model of the dipolariton}
In this section, we compare our model of the hybrid exciton and dipolariton with those applied in recent MoS$_2$ homobilayer experiments~\cite{Leisgang2020,Lorchat2021,Datta2022,Louca2023}. We model the dipolariton in an optical microcavity as a superposition of a cavity photon, two direct excitons (DXs) formed inside two spatially separated layers, and two indirect excitons (IXs) formed across the two layers with static out-of-plane electric dipole moments pointing in opposite directions, as illustrated in Fig.~\ref{fig:schematic}(a) of the main text. Setting $\hbar$ and the area to 1, the non-interacting Hamiltonian describing this system takes the form
\begin{align}
    \hat{H}_{0} = \sum_{\k} \Big[ \epsilon^{\text C}_{\k}\hat{c}^{\dagger}_{\k}\hat{c}_{\k} + \sum^{2}_{\eta=1}\epsilon^{\text{DX}}_{\k}\hat{x}^{\dagger}_{\k,\eta}\hat{x}_{\k,\eta} + \sum^{2}_{\eta=1}\epsilon^{\text{IX}}_{\k,\eta}\hat{y}^{\dagger}_{\k,\eta}\hat{y}_{\k,\eta} + \frac{\Omega}{2} \sum^{2}_{\eta=1} \left(\hat{x}^{\dagger}_{\k,\eta}\hat{c}_{\k}+\hat{c}^{\dagger}_{\k}\hat{x}_{\k,\eta}\right) + \frac{t}{2} \sum^{2}_{\eta=1}\left(\hat{x}^{\dagger}_{\k,\eta}\hat{y}_{\k,\eta}+\hat{y}^{\dagger}_{\k,\eta}\hat{x}_{\k,\eta}\right) \Big],
    \label{eq:Hamiltonian SM}
\end{align}
where we assume co-circular polarization for excitons and therefore restrict to a single spin-valley sector. Here, $\hat{c}_{\k}$ denotes the annihilation operator for a cavity photon. $\hat{x}_{\k,\eta}$ denotes the annihilation operator for a DX, where the index $\eta\in\{1,2\}$ labels layer and valley, with $\eta=1$ for the DX in the bottom layer and $K'$ valley, and $\eta=2$ in the top layer and $K$ valley. In our model, both DXs are taken to be B-direct excitons (B-DXs). Importantly, the models used in Refs.~\cite{Datta2022,Louca2023} do not fully account for the two B-DXs in the bilayer and thus fail to capture the dark exciton states resulting from the dark superposition of the DX$_1$ and DX$_2$. $\hat{y}_{\k,\eta}$ denotes the annihilation operator for an IX with an out-of-plane dipole along $\pm z$, where we define $+z$ to point from the bottom layer to the top layer. Accordingly, $\eta=1$ corresponds to the $+z$ dipole moment in the $K'$ valley and $\eta=2$ corresponds to the $-z$ dipole moment in the $K$ valley. The DXs are strongly coupled to the cavity photon via the Rabi splitting $\Omega$, while the IXs are coupled to the corresponding DX through the delocalization of the hole via the tunneling rate $t$.

The dispersions of the photon, DX, and IX modes take the forms
\begin{equation}
    \epsilon^\text{C}_{\k} = \frac{k^2}{2m_\text{C}} + \delta_\text{C}, \quad 
    \epsilon^{\text{DX}}_{\k} = \frac{k^2}{2m_\text{DX}}, \quad 
    \epsilon^{\text{IX}}_{\k,\eta}= \frac{k^2}{2m_\text{IX}}+\delta_\text{IX} \mp \Delta,
\end{equation}
respectively. $\delta_\text{C}$ denotes the photon-DX detuning and $\delta_\text{IX}$ denotes the IX-DX detuning. $\Delta$ denotes the Stark shift from an applied out-of-plane electric field, where the sign in front of $\Delta$ is taken to be $-$ for $\eta=1$ and $+$ for $\eta=2$. In the absence of light-matter coupling, the hole tunneling leads to DX$_{\eta}$-IX$_{\eta}$ hybridization for each $\eta$. This yields an upper (hX$^+$) and a lower (hX$^-$) hybrid exciton mode for each pair, whose energies are given by 
\begin{equation}
    E^{\pm}_{\k,\eta} = \frac{1}{2}\left( \epsilon^{\text{IX}}_{\k,\eta} + \epsilon^{\text{DX}}_{\k} \pm \sqrt{(\epsilon^{\text{IX}}_{\k,\eta}-\epsilon^{\text{DX}}_{\k})^{2} + t^{2}} \right).
\end{equation}
An applied out-of-plane electric field lifts the degeneracy of the IX energies, splitting the hX$^-$ mode into two branches associated with opposite dipole orientations~\cite{Leisgang2020,Lorchat2021}. For a sufficiently strong electric field (i.e., large $\Delta$), this effectively enables one to optically select a single dipole orientation in polariton-polariton scattering.

In MoS$_2$ homobilayers, the A-direct exciton (A-DX) is present below the hX$^{+}$ and hX$^{-}$ modes~\cite{Leisgang2020,Lorchat2021,Datta2022,Louca2023}. However, we do not include the A-DX in our model since it is essentially uncoupled from the physics of interest, i.e., the dipolariton interactions. While there is the possibility of coupling to the A-DX via the photon mode, we argue that this can be safely ignored because the maximum interactions between dipolaritons occur when the photon mode is positively detuned relative to the IX mode, which is well above the A-DX. It only becomes relevant when the photon energy approaches that of the A-DX, and this is far below the IX mode, where dipolar interactions no longer play a significant role.

\section{Exciton-exciton interaction potentials}
Here, we discuss in detail the pseudopotentials for the DX-DX, IX-IX, and DX-IX interactions. The potential operator is given by
\begin{align}
    \hat{V} &= \frac{1}{2}\sum_{\k\kprime\q}\sum^{2}_{\eta=1} V^{\eta\eta}_{\text{DX-DX}}(\q) \,\hat{x}^{\dagger}_{\k+\q,\eta}\hat{x}^{\dagger}_{\kprime-\q,\eta}\hat{x}_{\kprime,\eta}\hat{x}_{\k,\eta} + \frac{1}{2}\sum_{\k\kprime\q}\sum^{2}_{\eta=1} V^{\eta\eta}_{\text{IX-IX}}(\q) \,\hat{y}^{\dagger}_{\k+\q,\eta}\hat{y}^{\dagger}_{\kprime-\q,\eta}\hat{y}_{\kprime,\eta}\hat{y}_{\k,\eta} \notag \\
    &\quad + \sum_{\k\kprime\q}\sum^{2}_{\eta=1}V^{\eta\eta}_{\text{DX-IX}}(\q) \,\hat{x}^{\dagger}_{\k+\q,\eta}\hat{y}^{\dagger}_{\kprime-\q,\eta}\hat{y}_{\kprime,\eta}\hat{x}_{\k,\eta}.
    \label{eq:potential operator SM}
\end{align}
This is equivalent to Eq.~\eqref{eq:potential operator} in the main text, where we suppress superscripts on the potentials for simplicity. Note that intervalley (antiparallel) IX-IX interactions are omitted in Eq.~\eqref{eq:potential operator SM} because, within the regime of interest, the applied electric field makes one dipole orientation dominant in polariton-polariton scattering. Intervalley DX-DX and DX-IX interactions are likewise dropped, as their Born approximation vanishes at zero momentum.

We model intravalley DX-DX and DX-IX interactions as short-range soft-core potentials, and IX-IX interactions as dipole-dipole interactions at large distances with a short-range soft-core repulsion. In real space, these take the forms
\begin{subequations}
    \label{eq:V in real space SM}
\begin{align}
    V^{\eta\eta}_\text{DX-DX}(\mathbf{r}) &= V^{\eta\eta(\text{DX-DX})}_{0}\theta\left(a^{(\kappa)}_\text{X}-r\right), \\
    V^{\eta\eta}_\text{IX-IX}(\mathbf{r}) &= \frac{D^2}{\left(r^{(\kappa)}_{0}\right)^{3}}\theta\left(r^{(\kappa)}_{0}-r\right) + \frac{D^2}{r^{3}}\theta\left(r-r^{(\kappa)}_{0}\right), \\ 
    V^{\eta\eta}_\text{DX-IX}(\mathbf{r}) &= V^{\eta\eta(\text{DX-IX})}_{0}\theta\left(a^{(\kappa)}_\text{X}-r\right),
\end{align}
\end{subequations}
respectively, where $\theta(x)$ denotes the Heaviside step function. The potential range of the DX-DX and DX-IX interactions is set by the effective size of the exciton $a^{(\kappa)}_\text{X}=1/\sqrt{2\mu\varepsilon^{(\kappa)}_\text{X}}$, where $\varepsilon^{(\kappa)}_\text{X}$ is the DX binding energy for dielectric constant $\kappa$ and $\mu$ is the electron-hole reduced mass. $r^{(\kappa)}_{0}$ denotes the short-distance cutoff for the IX-IX interactions. We suppress the superscript for the dielectric constant $\kappa$ in the following unless it appears explicitly. The Fourier transforms of the potentials in Eq.~\eqref{eq:V in real space SM} are
\begin{subequations}
\begin{align}
    V^{\eta\eta}_\text{DX-DX}(\q) &= V^{\eta\eta(\text{DX-DX})}_{0}\frac{2\pi a_\text{X} J_{1}(qa_\text{X})}{q},  \\ 
    V^{\eta\eta}_\text{IX-IX}(\q) &= \frac{D^{2}}{r^{3}_{0}}\frac{2\pi r_{0} J_{1}(qr_{0})}{q} \notag \\ &\quad + D^2\bigg( -2\pi q + \pi qJ_{1}(qr_{0})[-2+\pi qr_{0}H_{0}(qr_{0})] + \frac{\pi}{r_{0}}J_{0}(qr_{0})[2+2q^2r^{2}_{0}-\pi q^2r^{2}_{0}H_{1}(qr_{0})] \biggl), \\ 
    V^{\eta\eta}_\text{DX-IX}(\q) &= V^{\eta\eta(\text{DX-IX})}_{0}\frac{2\pi a_\text{X} J_{1}(qa_\text{X})}{q},
\end{align}
    \label{eq:V in momentum space SM}
\end{subequations}
where $J_{n}(x)$ denotes the Bessel function of the first kind and $H_{n}(x)$ denotes the Struve function.

To determine the parameters of the potentials, i.e., $V_0$, $a_\text{X}$, and $r_0$, we stipulate that these interaction potentials reproduce the appropriate interaction properties of a more microscopic model, which explicitly accounts for the electrons and holes that form the bound excitons. Specifically, we will match the DX-DX, IX-IX, and DX-IX scattering within the Born approximation. Within our model, the Born approximations simply become
\begin{subequations}
    \label{eq:Born approximation SM}
\begin{align}
    V^{\eta\eta}_\text{DX-DX}(\q=0) &= V^{\eta\eta(\text{DX-DX})}_{0} \pi a^{2}_\text{X}, \label{eq:Born approximation for DX-DX SM} \\
    V^{\eta\eta}_\text{IX-IX}(\q=0) &= \frac{3\pi D^2}{r_{0}}, \label{eq:Born approximation for IX-IX SM} \\ 
    V^{\eta\eta}_\text{DX-IX}(\q=0) &= V^{\eta\eta(\text{DX-IX})}_{0} \pi a^{2}_\text{X}. \label{eq:Born approximation for DX-IX SM}
\end{align}
\end{subequations}
Calculating the Born approximation within a microscopic description is more complicated~\cite{Ciuti1998,Tassone1999}. We now go through the procedure in detail.

\subsection{Microscopic description of excitons}
To connect the effective interaction potentials to an underlying microscopic description of the excitons in terms of electrons and holes, in this and the next subsection, we consider the Hamiltonian
\begin{align}
    \hat{H}_\text{eh} &= \sum_{\k}\sum_{l,\xi}\left[(\epsilon_{\k}^\text{e} \hat{e}^\dag_{\k,l\xi}\hat{e}_{\k,l\xi}+\epsilon_{\k}^\text{h}\hat{h}^\dag_{\k,l\xi}\hat{h}_{\k,l\xi})\right] \notag \\ 
    &+ \frac{1}{2}\sum_{\substack{\k\k'\q}}\sum_{\substack{l,l'\\\xi,\xi'}} U_\q^{ll'}\left[ \hat{e}^\dag_{\k+\q,l\xi}\hat{e}^\dag_{\k'-\q,l'\xi'}\hat{e}_{\k',l'\xi'}\hat{e}_{\k,l\xi}+\hat{h}^\dag_{\k+\q,l\xi}\hat{h}^\dag_{\k'-\q,l'\xi'}\hat{h}_{\k',l'\xi'}\hat{h}_{\k,l\xi} -2\hat{e}^\dag_{\k+\q,l\xi}\hat{h}^\dag_{\k'-\q,l'\xi'}\hat{h}_{\k',l'\xi'}\hat{e}_{\k,l\xi} \right].
    \label{eq:microscopic Hamiltonian SM}
\end{align}
Here, $l=1,2$ indicates the layer and $\xi=K,K'$ the valley (see Fig.~\ref{fig:schematic}(a) in the main text). The dispersion is taken to be $\epsilon_\k^\text{e,h}=k^2/(2m_\text{e,h})$ where the electron and hole masses sum up to the exciton mass: $m_\text{X}=m_\text{e}+m_\text{h}$. The intra- and interlayer Rytova-Keldysh potentials are \cite{Asriyan2019,Semina2019}
\begin{subequations}
    \label{eq:Keldysh interaction SM}
\begin{align}
     U_\q^{11}&=U_\q^{22}= \frac{\pi}{\mu qa_{0}}\frac{1+q\rho_{0}(1-e^{-2qd})}{(1+q\rho_{0})^{2}-q^{2}\rho^{2}_{0}e^{-2qd}} \equiv U_{\q},
    \label{eq:intralayer Keldysh interaction SM}\\
      U^{12}_\q &= U^{21}_\q = \frac{\pi}{\mu qa_{0}}\frac{e^{-qd}}{(1+q\rho_{0})^{2}-q^{2}\rho^{2}_{0}e^{-2qd}} \equiv U^{d}_{\q},
    \label{eq:interlayer Keldysh interaction SM}
\end{align}
\end{subequations}
respectively, where $\mu=m_\text{e}m_\text{h}/(m_\text{e}+m_\text{h})$ is the reduced mass. Here, the dielectric screening length takes the form $\rho^{(\kappa)}_{0} = 2\pi\chi_\text{2D}/\kappa$, where $\chi_\text{2D}$ is the 2D polarizability of the layers. The effective 2D Bohr radius is given by $a^{(\kappa)}_{0} = \kappa/(2\mu e^{2})$, with $e$ denoting the elementary charge. We suppress the superscript for $\kappa$ unless otherwise noted. Importantly, the intralayer Rytova-Keldysh potential in Eq.~\eqref{eq:intralayer Keldysh interaction SM} differs from the monolayer Rytova-Keldysh potential~\cite{Rytova1967,Keldysh1979} due to the presence of the second layer, which provides additional screening of the electron-hole interactions. Taking the limit $\rho_{0} \to 0$, Eqs.~\eqref{eq:intralayer Keldysh interaction SM} and \eqref{eq:interlayer Keldysh interaction SM}, respectively, reduce to the usual intra- and interlayer Coulomb potentials.

In writing the microscopic Hamiltonian in Eq.~\eqref{eq:microscopic Hamiltonian SM}, we have neglected the hole tunneling and the light-matter coupling, since we use this Hamiltonian to independently evaluate the Born approximation for DX-DX, IX-IX, and DX-IX scattering within the microscopic description. Matching the Born approximation for our pseudopotentials in Eq.~\eqref{eq:Born approximation SM} with that obtained from the microscopic description ensures that the pseudopotentials capture the correct upper bound of the exciton $T$ matrix. We then apply the pseudopotentials in polariton-polariton scattering while assuming that neither the light-matter coupling nor the hole tunneling affects the exciton-exciton interactions. This is reasonable since both the hole tunneling and Rabi splitting are smaller than the exciton binding energies. As such, we do not consider their influence in our microscopic Hamiltonian. Instead, we introduce these terms into the exciton Hamiltonian in Eq.~\eqref{eq:Hamiltonian SM} in a manner consistent with Refs.~\cite{Maslova2024,Byrnes2014,Nalitov2014b}, although Ref.~\cite{Maslova2024} went further and included tunneling into their calculation of the interaction matrix elements.

\begin{figure}[t]
    \centering
    \begin{minipage}{0.99\textwidth}
    \centering
    \includegraphics[width=\textwidth]{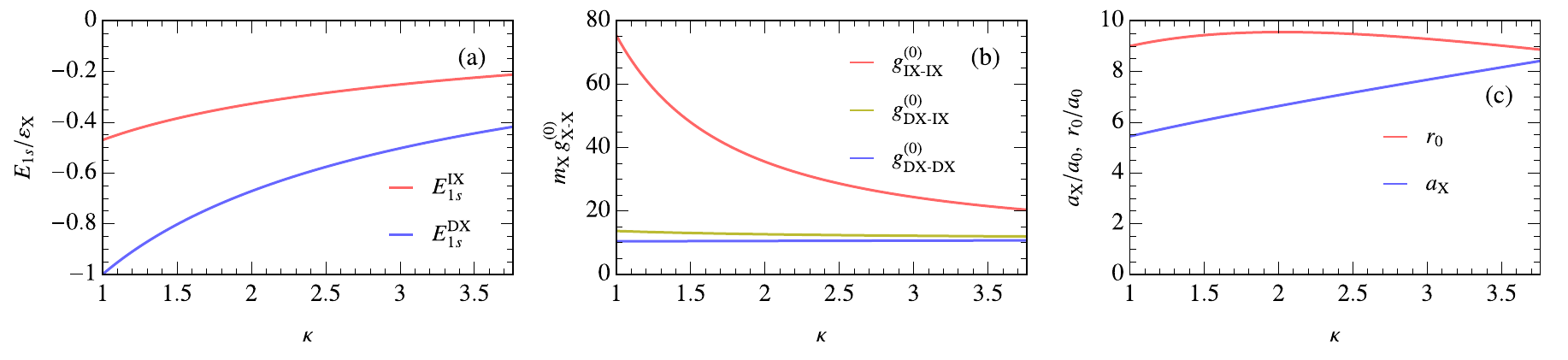}
    \end{minipage}
    \caption{(a) Energy of the ground-state DX (blue) and IX (red) as a function of dielectric constant. We measure the energy from the corresponding electron-hole continuum. (b) Interaction constant for intravalley DX-DX (blue), IX-IX (red), and DX-IX (yellow) scattering obtained from the Born approximation. We suppress the index $\eta$ for simplicity. (c) Effective size of the exciton $a_\text{X}$ for short-range interactions, and short-distance cutoff $r_{0}$ for dipolar interactions. In all panels, we take equal masses for the electron and hole, and we use $\rho_{0}/a_{0} = 40$ and $d/a_{0} = 6$, with the reference scales $\rho_{0}$, $a_{0}$, and $\varepsilon_\text{X}$ fixed by taking $\kappa=1$.}
    \label{fig:microscopic description SM}
\end{figure}

In order to describe direct and indirect excitons, we consider the most general wave functions for a direct exciton and an indirect exciton at zero center-of-mass momentum
\begin{subequations}
\begin{align}
    \ket{\Phi_{\text{DX},l\xi}} & =\sum_{\k} \phi_{\k} \hat{e}^{\dag}_{\k,l\xi}\hat{h}^{\dag}_{-\k,l\xi}\ket{0}, \\
    \ket{\Psi_{\text{IX},l\xi}} & =\sum_{\k} \psi_{\k} \hat{e}^{\dag}_{\k,l\xi}\hat{h}^{\dag}_{-\k,\bar{l}\xi}\ket{0},
\end{align}
\end{subequations}
where $\ket{0}$ denotes the vacuum state. We use $\bar{l}$ to denote the index opposite to $l$, and we only consider excitons formed by the electron and hole residing in the same valley. By considering the \sch equations $\hat{H}_\text{eh}\ket{\Phi_{\text{DX},l\xi}}=E\ket{\Phi_{\text{DX},l\xi}}$ and $\hat{H}_\text{eh}\ket{\Psi_{\text{IX},l\xi}}=E\ket{\Psi_{\text{IX},l\xi}}$, we find that the wave functions satisfy
\begin{subequations}
\label{eq:sch}
\begin{align}
    E\phi_{\k} &= \bar\epsilon_{\k}\phi_{\k} - \sum_{\kprime}U_{\k-\kprime}\phi_{\kprime},
    \label{eq:schrodinger equationDX}\\
    E\psi_{\k} &= \bar\epsilon_{\k}\psi_{\k} - \sum_{\kprime}U^d_{\k-\kprime}\psi_{\kprime},
    \label{eq:schrodinger equationIX}
\end{align}
\end{subequations}
where we have defined the total kinetic energy $\bar\epsilon_\k=\epsilon_\k^\text{e}+\epsilon_\k^\text{h}$. For a finite screening length $\rho_{0}$, both Eq.~\eqref{eq:schrodinger equationDX} and Eq.~\eqref{eq:schrodinger equationIX} are solved numerically to obtain the DX and IX ground-state wave functions $\Phi_{\k}$ and $\Psi_{\k}$ and the corresponding energies $E_\text{DX}=-\varepsilon^{(\kappa)}_\text{X}$ and $E_\text{IX}$. In the Coulomb limit ($\rho_{0} \to 0$), the DX ground-state wave function is well known to take the analytic form
\begin{align}
    \Phi_\k=\frac{\sqrt{8\pi}a_0}{(1+k^2a_0^2)^{3/2}}.
\end{align}
In Fig.~\ref{fig:microscopic description SM}(a), we show the DX and IX binding energies as a function of dielectric constant at fixed interlayer separation. We observe that the presence of the surrounding dielectric environment reduces the DX and IX binding energies. Note that, within this section, we measure both energies from their respective electron-hole continuum, which is convenient since we will not be directly comparing their energies.

\subsection{Born approximation for exciton-exciton interactions}

We now consider the Born approximation for the DX-DX, IX-IX, and DX-IX scattering. The idea is to evaluate the interaction energy shift by approximating the exact interacting two-exciton state by that of two uncorrelated excitons. As such, the Born approximation can be viewed as a variational ansatz that provides an upper bound for the exciton interaction strength~\cite{Li2021a}.

Specifically, for the case of direct excitons in the ground state, we consider the corresponding operator $\hat{x}_{\eta}$ with $\eta \in \{1,2\}$, where the index $\eta$ accounts for the layer and valley
\begin{equation}
    \hat{x}_{\0,1} = \sum_{\k} \Phi_{\k} \hat{e}_{\k,1K'}\hat{h}_{-\k,1K'}, \quad \hat{x}_{\0,2} = \sum_{\k} \Phi_{\k}\hat{e}_{\k,2K}\hat{h}_{-\k,2K}.
\end{equation}
This is the microscopic equivalent of the DX operator $\hat{x}_{\eta}$ in the main text. We then follow Ref.~\cite{Levinsen2019a} and evaluate the intravalley DX-DX Born approximation as
\begin{align}
    g^{\eta\eta(0)}_\text{DX-DX} &= \frac{1}{2}\bra{0}\hat{x}_{\0,\eta}\hat{x}_{\0,\eta}(\hat{H}_\text{eh}-2E_\text{DX})\hat{x}^{\dagger}_{\0,\eta}\hat{x}^{\dagger}_{\0,\eta}\ket{0} \notag \\ 
    &= 2\sum_{\k}({\bar\epsilon}_{\k}-E_\text{DX})\Phi^{4}_{\k} - 2\sum_{\k\kprime}U_{\k-\kprime}\Phi^{2}_{\k}\Phi^{2}_{\kprime},
    \label{eq:microscopic Born approximation for DX-DX SM}
\end{align}
which matches that obtained in Ref.~\cite{Pico2025}, and further reduces to $g^{\eta\eta(0)}_\text{DX-DX} \simeq 6\varepsilon_\text{X}a_0^2$ in the Coulomb limit in Refs.~\cite{Ciuti1998,Tassone1999}. Turning instead to the case of indirect excitons, we define the corresponding operator $\hat{y}_{\eta}$ with $\eta \in \{1,2\}$, where the index $\eta$ accounts for the dipole orientation and valley. We specifically define the indirect exciton operator as
\begin{equation}
    \hat{y}_{\0,1} = \sum_{\k} \Psi_{\k} \hat{e}_{\k,1K'}\hat{h}_{-\k,2K'}, \quad \hat{y}_{\0,2} = \sum_\k \Psi_\k \hat{e}_{\k,2K}\hat{h}_{-\k,1K}.
\end{equation}
Remarkably, we find that the Born approximation for the intravalley (parallel) IX-IX scattering takes a functional form that is very similar to that above
\begin{align}
    g^{\eta\eta(0)}_\text{IX-IX} &= \frac{1}{2}\bra{0}\hat{y}_{\0,\eta}\hat{y}_{\0,\eta}(\hat{H}_\text{eh}-2E_\text{IX})\hat{y}^{\dagger}_{\0,\eta}\hat{y}^{\dagger}_{\0,\eta}\ket{0} \notag \\ 
    &= 2\sum_{\k}({\bar\epsilon}_{\k}-E_\text{IX})\Psi^{4}_{\k} - 2\sum_{\k\kprime}U_{\k-\kprime}\Psi^{2}_{\k}\Psi^{2}_{\kprime}+ \frac{2\pi d}{\mu a_{0}}.
    \label{eq:microscopic Born approximation for IX-IX SM}
\end{align}
The first and the second terms in Eq.~\eqref{eq:microscopic Born approximation for IX-IX SM} result from the exchange interactions while the third term corresponds to a Hartree term. Likewise, the Born approximation for the intravalley DX-IX scattering takes the form
\begin{align}
    g^{\eta\eta(0)}_\text{DX-IX} &= \bra{0}\hat{y}_{\0,\eta}\hat{x}_{\0,\eta}(\hat{H}_\text{eh}-E_\text{DX}-E_\text{IX})\hat{x}^{\dagger}_{\0,\eta}\hat{y}^{\dagger}_{\0,\eta}\ket{0} \notag \\ 
    &= \sum_{\k}(2{\bar\epsilon}_{\k}-E_\text{DX}-E_\text{IX})\Phi_{\k}^{2}\Psi^{2}_{\k} - \sum_{\k\kprime}(U_{\k-\kprime}+U^{d}_{\k-\kprime})\Phi^{2}_{\k}\Psi^{2}_{\kprime},
    \label{eq:microscopic Born approximation for DX-IX SM}
\end{align}
for $\eta=1,2$. 

In the case of the intervalley DX-DX, IX-IX, and DX-IX scattering, the constituent electrons and holes forming the excitons are distinguishable; therefore, only the Hartree term contributes to the Born approximation. In the limit of zero momentum scattering, we have
\begin{align}
    g^{\eta\bar{\eta}(0)}_\text{DX-DX} &= \bra{0}\hat{x}_{\0,\bar{\eta}}\hat{x}_{\0,\eta}(\hat{H}_\text{eh}-2E_\text{DX})\hat{x}^{\dagger}_{\0,\eta}\hat{x}^{\dagger}_{\0,\bar{\eta}}\ket{0} \\ \notag 
    &= 0, \\[0.2cm] 
    g^{\eta\bar{\eta}(0)}_\text{IX-IX} &= \bra{0}\hat{y}_{\0,\bar{\eta}}\hat{y}_{\0,\eta}(\hat{H}_\text{eh}-2E_\text{IX})\hat{y}^{\dagger}_{\0,\eta}\hat{y}^{\dagger}_{\0,\bar{\eta}}\ket{0} \notag \\ 
    &= -\frac{2\pi d}{\mu a_{0}}, \\[0.2cm]
    g^{\eta\bar{\eta}(0)}_\text{DX-IX} &= \bra{0}\hat{y}_{\0,\bar{\eta}}\hat{x}_{\0,\eta}(\hat{H}_\text{eh}-E_\text{DX}-E_\text{IX})\hat{x}^{\dagger}_{\0,\eta}\hat{y}^{\dagger}_{\0,\bar{\eta}}\ket{0} \notag \\ 
    &= 0,
\end{align}
where $\bar{\eta}$ denotes the index opposite to $\eta$. While we do not consider intervalley (antiparallel) IX-IX interactions in the exciton potential operator in Eq.~\eqref{eq:potential operator SM}, we present the microscopic result for the sake of generality.

In Fig.~\ref{fig:microscopic description SM}(b), we show the interaction constant obtained from the Born approximation for the intravalley DX-DX, IX-IX, and DX-IX scattering as a function of dielectric constant at fixed interlayer separation. The key feature is that the IX-IX scattering can be significantly enhanced towards the vacuum limit ($\kappa \to 1$), since the effective 2D Bohr radius becomes smaller in this regime and the Hartree term in Eq.~\eqref{eq:microscopic Born approximation for IX-IX SM} enhances scattering between IXs without changing the interlayer separation. This is in stark contrast to the DX-DX scattering, whose interaction strength slightly decreases as the dielectric constant of the surrounding environment decreases. Although the interaction constant for the DX-IX scattering slightly increases towards the vacuum limit, this term provides only a marginal contribution compared to IX-IX interactions in dipolariton scattering, as discussed below.

We are now in a position to fix the parameters of the pseudopotentials in Eq.~\eqref{eq:V in real space SM}. For simplicity, we take equal electron and hole masses and thus $\mu=m_\text{X}/4$. In the case of the repulsive DX-DX and DX-IX potentials, we take the range $a_\text{X}$ to equal the effective size of the exciton. Hence, to match the Born approximation for the pseudopotentials in Eqs.~\eqref{eq:Born approximation for DX-DX SM} and \eqref{eq:Born approximation for DX-IX SM} with the microscopic Born approximation in Eqs.~\eqref{eq:microscopic Born approximation for DX-DX SM} and \eqref{eq:microscopic Born approximation for DX-IX SM}, we have
\begin{align}
    V^{\eta\eta(\text{DX-DX})}_{0} &= \frac{g^{\eta\eta(0)}_\text{DX-DX}}{\pi \left(a^{(\kappa)}_\text{X}\right)^{2}}, \\
    V^{\eta\eta(\text{DX-IX})}_{0} &= \frac{g^{\eta\eta(0)}_\text{DX-IX}}{\pi \left(a^{(\kappa)}_\text{X}\right)^{2}}.
\end{align}
Similarly, we define the short-distance cutoff $r_{0}$ for the dipolar interactions such that the Born approximation for $V^{\eta\eta}_\text{IX-IX}$ in Eq.~\eqref{eq:Born approximation for IX-IX SM} equals that of the microscopic description in Eq.~\eqref{eq:microscopic Born approximation for IX-IX SM}. Therefore, we identify $r_{0}$ as
\begin{equation}
    r^{(\kappa)}_{0} = \frac{3\pi D^{2}}{g^{\eta\eta(0)}_\text{IX-IX}} = \frac{3\pi d^{2}}{2\mu a^{(\kappa)}_{0}g^{\eta\eta(0)}_\text{IX-IX}},
\end{equation}
where we have used $D^{2} = d^{2}/(2\mu a^{(\kappa)}_{0})$. In Fig.~\ref{fig:microscopic description SM}(c), we show the effective size of the exciton and the short-distance cutoff as a function of dielectric constant. As expected from the DX energy in Fig.~\ref{fig:microscopic description SM}(a), the effective size of the exciton, and hence the range of the soft-core potential, increases for larger dielectric constants since the DX is weakly bound. On the other hand, the short-distance cut-off for the dipolar interactions is relatively insensitive to the dielectric constant within the regime we consider, and it remains slightly larger than the interlayer separation.

\section{Scattering of dipolaritons}
In this section, we provide additional details on the derivation of the $T$-matrix equation used for the calculations presented in the main text, extending the formalism developed in Ref.~\cite{Bleu2020} for polariton-polariton scattering with the contact interaction. Therefore, we now consider the exact scattering in our system, calculated within the Hamiltonian in Eq.~\eqref{eq:Hamiltonian SM}. We define the two-particle state with zero center-of-mass momentum
\begin{equation}
    \ket{A,B,\k} = \hat{A}^{\dagger}_{\k}\hat{B}^{\dagger}_{-\k}\ket{0}.
    \label{eq:two-particle states SM}
\end{equation}
Here, ${\hat{A}}$ and $\hat{B}$ correspond to the photon operator $\hat{c}$, the DX$_1$ operator $\hat{x}_{1}$, the DX$_2$ operator $\hat{x}_{2}$, the IX$_1$ operator $\hat{y}_{1}$, or the IX$_2$ operator $\hat{y}_{2}$, which follow the bosonic commutation relations
\begin{equation}
    \left[ \hat{A}_{\k},\hat{B}^{\dagger}_{\k} \right] = \delta_{AB}\delta_{\k,\k'}.
\end{equation}
The scalar product of the two-particle states reads
\begin{align}
    \bra{A,B,\k'}\ket{C,D,\k} &= \bra{0} \hat{B}_{-\k'}\hat{A}_{\k'}\hat{C}^{\dagger}_{\k}\hat{D}^{\dagger}_{-\k} \ket{0} \notag \\
    &= \delta_{AC}\delta_{BD}\delta_{\k',\k} + \delta_{AD}\delta_{BC}\delta_{\k',-\k}.
    \label{eq:scalar product SM}
\end{align}
The identity operator follows the completeness relation
\begin{equation}
    \mathbb{1} = \frac{1}{2}\sum_{\q}\sum_{A,B=\{c,x_{1},x_{2},y_{1},y_{2}\}} \ket{A,B,\q}\bra{A,B,\q},
    \label{eq:completeness relation SM}
\end{equation}
where the factor of $1/2$ is required to avoid double counting. The identity operator satisfies the property
\begin{align}
    \mathbb{1}\ket{C,D,\k} &= \frac{1}{2}\sum_{\q}\sum_{A,B=\{c,x_{1},x_{2},y_{1},y_{2}\}} \ket{A,B,\q}\bra{A,B,\q}\ket{C,D,\k} \notag \\
    &= \frac{1}{2}\sum_{\q}\sum_{A,B=\{c,x_{1},x_{2},y_{1},y_{2}\}} (\delta_{AC}\delta_{BD}\delta_{\q,\k} + \delta_{AD}\delta_{BC}\delta_{\q,-\k}) \hat{A}^{\dagger}_{\q}\hat{B}^{\dagger}_{-\q}\ket{0} \notag \\
    &= \frac{1}{2}\sum_{\q}(\hat{C}^{\dagger}_{\q}\hat{D}^{\dagger}_{-\q}\ket{0}\delta_{\q,\k} + \hat{D}^{\dagger}_{\q}\hat{C}^{\dagger}_{-\q}\ket{0}\delta_{\q,-\k}) \notag \\
    &= \ket{C,D,\k}
\end{align}
as expected.

As the interacting states relevant for the $T$-matrix equation, we introduce the index notation $\ket{i,\k}$ for the two-particle states in Eq.~\eqref{eq:two-particle states SM}. The interacting two-particle states for $i=1,\cdots,8$ correspond to
\begin{align}
    &\ket{1,\k} = \hat{x}^{\dagger}_{\k,1}\hat{x}^{\dagger}_{-\k,1}\ket{0}, \quad
    \ket{2,\k} = \hat{x}^{\dagger}_{\k,2}\hat{x}^{\dagger}_{-\k,2}\ket{0}, \quad
    \ket{3,\k} = \hat{y}^{\dagger}_{\k,1}\hat{y}^{\dagger}_{-\k,1}\ket{0}, \quad 
    \ket{4,\k} = \hat{y}^{\dagger}_{\k,2}\hat{y}^{\dagger}_{-\k,2}\ket{0}, \notag \\[0.2cm] 
    &\ket{5,\k} = \frac{1}{\sqrt{2}} \left( \hat{x}^{\dagger}_{\k,1}\hat{y}^{\dagger}_{-\k,1} + \hat{y}^{\dagger}_{\k,1}\hat{x}^{\dagger}_{-\k,1} \right) \ket{0}, \quad 
    \ket{6,\k} = \frac{1}{\sqrt{2}} \left( \hat{x}^{\dagger}_{\k,2}\hat{y}^{\dagger}_{-\k,2} + \hat{y}^{\dagger}_{\k,2}\hat{x}^{\dagger}_{-\k,2} \right) \ket{0}, \notag \\[0.2cm]
    &\ket{7,\k} = \frac{1}{\sqrt{2}} \left( \hat{x}^{\dagger}_{\k,1}\hat{y}^{\dagger}_{-\k,1} - \hat{y}^{\dagger}_{\k,1}\hat{x}^{\dagger}_{-\k,1} \right) \ket{0}, \quad 
    \ket{8,\k} = \frac{1}{\sqrt{2}} \left( \hat{x}^{\dagger}_{\k,2}\hat{y}^{\dagger}_{-\k,2} - \hat{y}^{\dagger}_{\k,2}\hat{x}^{\dagger}_{-\k,2} \right) \ket{0},
    \label{eq:interacting two-particle states SM}
\end{align}
while the remaining non-interacting two-body states correspond to $i=9,\cdots,25$. In writing Eq.~\eqref{eq:interacting two-particle states SM} for $i=5,6$, we have expressed the two-particle states by taking the symmetric combinations of the corresponding excitons, which effectively bosonize the two-particle states. Since their antisymmetric combinations with $i=7,8$ behave as fermionic states, they decouple from the $T$-matrix equation for even partial waves. On this basis, the identity operator in Eq.~\eqref{eq:completeness relation SM} takes the form
\begin{equation}
    \mathbb{1} = \frac{1}{2}\sum_{\q}\sum^{25}_{i=1} \ket{i,\q}\bra{i,\q}.
    \label{eq:completeness relation 2 SM}
\end{equation}

We now consider the momentum space representation of the Lippmann-Schwinger equation
\begin{equation}
    \bra{i,\k'}\hat{T}\ket{j,\k} = \bra{i,\k'}\hat{V}\ket{j,\k} + \bra{i,\k'}\hat{V}\hat{G}\hat{T}\ket{j,\k}.
    \label{eq:Lippmann-Schwinger equation SM}
\end{equation}
Here, the Green's operator is defined as
\begin{align}
    \hat{G}(E) = \frac1{(E+i0)\mathbb{1}-\hat{H}_{0}}.
    \label{eq:Green's operator SM}
\end{align}
The matrix element of the potential operator for the two-particle state in Eq.~\eqref{eq:interacting two-particle states SM} takes the form
\begin{equation}
    \bra{i,\k'}\hat{V}\ket{j,\k} = V_{ij}(\k'-\k)\delta_{ij} \pm V_{ij}(\k'+\k)\delta_{ij},
    \label{eq:potential matrix element SM}
\end{equation}
where the $+$ sign corresponds to the bosonic states with $i=1,\cdots,6$, while the $-$ sign corresponds to the fermionic states with $i=7,8$. Using the correspondence between index notation for the two-particle state and the exciton interaction potentials in the potential operator, we define $V_{11(22)}= V^{\eta\eta=11(22)}_\text{DX-DX}$, $V_{33(44)}=V^{\eta\eta=11(22)}_\text{IX-IX}$, $V_{55(66)}= V^{\eta\eta=11(22)}_\text{DX-IX}$, $V_{77(88)}= V^{\eta\eta=11(22)}_\text{DX-IX}$, and $V_{ij} = 0$ when $i=9,\cdots,25$ or $j=9,\cdots,25$. This implies that the $T$-matrix element $\bra{i,\k'}\hat{T}\ket{j,\k} = 0$ when $i=9,\cdots,25$ or $j=9,\cdots,25$. Inserting the identity operator in Eq.~\eqref{eq:completeness relation 2 SM} into Eq.~\eqref{eq:Lippmann-Schwinger equation SM}, we obtain
\begin{align}
    &\bra{i,\k'}\hat{V}\hat{G}\hat{T}\ket{j,\k} = \bra{i,\k'}\hat{V}\mathbb{1}\hat{G}\mathbb{1}\hat{T}\ket{j,\k} \notag \\ 
    &= \frac{1}{4}\sum_{\q'}\sum^{25}_{m=1}\sum_{\q}\sum^{25}_{n=1}\bra{i,\k'}\hat{V}\ket{m,\q'}\bra{m,\q'}\hat{G}\ket{n,\q}\bra{n,\q}\hat{T}\ket{j,\k} \notag \\
    &= \frac{1}{4}\sum_{\q'}\sum_{\q}\sum^{8}_{n=1}    \bra{i,\k'}\hat{V}\ket{i,\q'}\bra{i,\q'}\hat{G}\ket{n,\q}\bra{n,\q}\hat{T}\ket{j,\k} \notag \\
    &= \frac{1}{4}\sum_{\q'}\sum_{\q}\sum^{6}_{n=1}    \bra{i,\k'}\hat{V}\ket{i,\q'}G_{in}(\q,E)\bra{n,\q}\hat{T}\ket{j,\k}(\delta_{\q',\q} + \delta_{\q',-\q}) \notag \\ 
    &\quad + \frac{1}{4}\sum_{\q'}\sum_{\q}\sum^{8}_{n=7}    \bra{i,\k'}\hat{V}\ket{i,\q'}G_{in}(\q,E)\bra{n,\q}\hat{T}\ket{j,\k}(\delta_{\q',\q} - \delta_{\q',-\q}) \notag \\
    &= \frac{1}{2}\sum_{\q}\sum^{8}_{n=1}    \bra{i,\k'}\hat{V}\ket{i,\q}G_{in}(\q,E)\bra{n,\q}\hat{T}\ket{j,\k}.
\end{align}
In the third line, we have applied Eq.~\eqref{eq:potential matrix element SM}, and in the fourth line, we have defined the propagator as $\bra{i,\q'}\hat{G}(E)\ket{n,\q} = G_{in}(\q,E)(\delta_{\q',\q} \pm \delta_{\q',-\q})$, where the $+$ sign corresponds to $i,n=1,\cdots,6$ and the $-$ sign to $i,n=7,8$. In the fifth line, we have applied the relation $\ket{n,\q} = \ket{n,-\q}$ when $n=1,\cdots,6$ and $\ket{n,\q} = -\ket{n,-\q}$ when $n=7,8$. We discuss the details of the propagator below. Thus, we obtain the momentum-space representation of the Lippmann-Schwinger equation
\begin{equation}
    \bra{i,\k'}\hat{T}\ket{j,\k} = \bra{i,\k'}\hat{V}\ket{j,\k} + \frac{1}{2}\sum_{\q}\sum^{8}_{n=1}    \bra{i,\k'}\hat{V}\ket{i,\q}G_{in}(\q,E)\bra{n,\q}\hat{T}\ket{j,\k}.
    \label{eq:Lippmann-Schwinger equation 2 SM}
\end{equation}
In the case of scattering with rotationally symmetric potentials and zero center-of-mass momentum, Eq.~\eqref{eq:Lippmann-Schwinger equation 2 SM} can be reduced to the $l$-wave scattering integral equation~\cite{Dickhoff2008}
\begin{equation}
    \bra{i,k'l}\hat{T}\ket{j,kl} = \bra{i,k'l}\hat{V}\ket{j,kl} + \frac{1}{2}\sum^{8}_{n=1}\int^{\infty}_{0}\frac{q\,dq}{2\pi} \bra{i,k'l}\hat{V}\ket{i,ql}G_{in}(q,E)\bra{n,ql}\hat{T}\ket{j,kl},
    \label{eq:partial wave Lippmann-Schwinger equation equation SM}
\end{equation}
where $l$ denotes the angular momentum quantum number.

To obtain the scattering properties, we need the matrix elements of the interaction potentials. Since we have rotationally symmetric potentials and zero center-of-mass momentum, these can be expanded into partial waves as
\begin{equation}
    V(\k'\pm\k) = \sum^{\infty}_{l=0}(2-\delta_{l0})\cos(l\theta_{\k'\k})V^{(l)}_{\pm}(k',k),
    \label{eq:partial wave decomposition of V SM}
\end{equation}
with $\theta_{\k'\k}$ denoting the relative angle between $\k'$ and $\k$. $V$ denotes the general form of the interaction potential, which can correspond to $V_\text{DX-DX}$, $V_\text{IX-IX}$, or $V_\text{DX-IX}$. The $l$-wave projection of the interaction potentials can be found by inverting Eq.~\eqref{eq:partial wave decomposition of V SM}, which takes the form
\begin{equation}
    V^{(l)}_{\pm}(k',k) = \int^{2\pi}_{0}\frac{d\theta_{\k'\k}}{2\pi}\cos(l\theta_{\k'\k})V(\k'\pm\k),
    \label{eq:partial wave projection of V SM}
\end{equation}
where $V^{(l)}_{+}$ and $V^{(l)}_{-}$ are related via
\begin{align}
     V^{(l)}_{+}(k',k) =
    \begin{dcases}
    V^{(l)}_{-}(k',k), \quad &\text{$l$ is even} \\[0.2cm]
    -V^{(l)}_{-}(k',k), \quad &\text{$l$ is odd}.
    \end{dcases}
    \label{eq:potential matrix element def}
\end{align}
Therefore, the matrix element in Eq.~\eqref{eq:potential matrix element SM} reduces to
\begin{equation}
    \bra{i,k'l}\hat{V}\ket{j,kl} = 
    \begin{dcases}
    2V^{(l)}_{ij}(k',k)\delta_{ij}, \quad &\text{$l$ is even} \\[0.2cm]
    0, \quad &\text{$l$ is odd}
    \end{dcases}
    \quad \text{and $i,j \in \{1,\cdots,6\}$},
    \label{eq:sym partial wave potential element SM}
\end{equation}
for the symmetric states, and
\begin{equation}
    \bra{i,k'l}\hat{V}\ket{j,kl} = 
    \begin{dcases}
    0, \quad &\text{$l$ is even} \\[0.2cm]
    2V^{(l)}_{ij}(k',k)\delta_{ij}, \quad &\text{$l$ is odd}
    \end{dcases}
    \quad \text{and $i,j \in \{7,8\}$},
    \label{eq:asym partial wave potential element SM}
\end{equation}
for the antisymmetric states. We further define the $l$-wave $T$ matrix as
\begin{equation}
    \bra{i,k'l}\hat{T}\ket{j,kl} = 2T^{(l)}_{ij}(k',k;E),
    \label{eq:partial wave T-matrix element SM}
\end{equation}
where the factor of 2 accounts for the direct and exchange contributions to scattering of identical particles. Using Eqs.~\eqref{eq:sym partial wave potential element SM}, \eqref{eq:asym partial wave potential element SM}, and \eqref{eq:partial wave T-matrix element SM} in Eq.~\eqref{eq:partial wave Lippmann-Schwinger equation equation SM}, we obtain the coupled integral equations
\begin{equation}
    T^{(l)}_{ij}(k',k;E) = V^{(l)}_{ij}(k',k)\delta_{ij} + \sum^{6}_{n=1}\int^{\infty}_{0}\frac{q\,dq}{2\pi} V^{(l)}_{ii}(k',q)G_{in}(q,E)T^{(l)}_{nj}(q,k;E), \quad \text{$i,j \in \{1,\cdots,6\}$ and $l$ is even},
    \label{eq:even partial wave scattering integral equation SM}
\end{equation}
with $V_{11(22)}= V^{\eta\eta=11(22)}_\text{DX-DX}$, $V_{33(44)}=V^{\eta\eta=11(22)}_\text{IX-IX}$, and $V_{55(66)}= V^{\eta\eta=11(22)}_\text{DX-IX}$, and
\begin{equation}
    T^{(l)}_{ij}(k',k;E) = V^{(l)}_{ij}(k',k)\delta_{ij} + \sum^{8}_{n=7}\int^{\infty}_{0}\frac{q\,dq}{2\pi} V^{(l)}_{ii}(k',q)G_{in}(q,E)T^{(l)}_{nj}(q,k;E), \quad \text{$i,j \in \{7,8\}$ and $l$ is odd},
    \label{eq:odd partial wave scattering integral equation SM}
\end{equation}
with $V_{77(88)}= V^{\eta\eta=11(22)}_\text{DX-IX}$. The $s$-wave $(l=0)$ equation in Eq.~\eqref{eq:even partial wave scattering integral equation SM} corresponds to Eq.~\eqref{eq:scattering integral equation} of the main text.

We now define the propagators relevant for the scattering integral equation in Eq.~\eqref{eq:even partial wave scattering integral equation SM}. Denoting the propagators for dipolariton scattering as $G^{P}_{in}(\q,E) = \bra{i,\q}\hat{G}(E)\ket{n,\q}$, these take the form
\begin{equation}
    G^{P}_{in}(\q,E) = \sum^{6}_{m_{1},m_{2}=1} \frac{\bra{i,\q} \ket{P_{m_1},P_{m_2},\q}\bra{P_{m_1},P_{m_2},\q}\ket{n,\q}}{E-E^{(m_1)}_{\q}-E^{(m_2)}_{\q}+i0},
\end{equation}
for $i,n=1,\cdots,6$, where $\ket{P_{m_1},P_{m_2},\q} = \hat P_{m_1,\q}^{\dagger}\hat P_{m_2,-\q}^{\dagger}\ket{0}$. Here, $\bra{i,\q}\ket{P_{m_1},P_{m_2},\q}$ corresponds to the product of excitonic transformation coefficients for two propagating polaritons $P_{m_1}$ and $P_{m_2}$ when the outgoing interacting two-particle state is $\ket{i,\q}$ in Eq.~\eqref{eq:interacting two-particle states SM}. More specifically, this takes the form
\begin{equation}
    \bra{i,\q}\ket{P_{m_1},P_{m_2},\q} =
    \begin{dcases}
    X^{(m_1)}_{\q,1}X^{(m_2)}_{\q,1}, \quad &\text{for} \quad i = 1, \\[0.2cm]
    X^{(m_1)}_{\q,2}X^{(m_2)}_{\q,2}, \quad &\text{for} \quad i = 2, \\[0.2cm]
    Y^{(m_1)}_{\q,1}Y^{(m_2)}_{\q,1}, \quad &\text{for} \quad i = 3, \\[0.2cm] 
    Y^{(m_1)}_{\q,2}Y^{(m_2)}_{\q,2}, \quad &\text{for} \quad i = 4, \\[0.2cm] 
    \frac{1}{\sqrt{2}} \left( X^{(m_1)}_{\q,1}Y^{(m_2)}_{\q,1}+Y^{(m_1)}_{\q,1}X^{(m_2)}_{\q,1} \right), \quad &\text{for} \quad i = 5, \\[0.2cm]
    \frac{1}{\sqrt{2}} \left( X^{(m_1)}_{\q,2}Y^{(m_2)}_{\q,2}+Y^{(m_1)}_{\q,2}X^{(m_2)}_{\q,2} \right), \quad &\text{for} \quad i = 6.
    \end{dcases}
    \label{eq:excitonic transformation coefficients SM}
\end{equation}
The polariton dispersions and the transformation coefficients satisfy Eqs.~\eqref{eq:eigenvalues} and \eqref{eq:linear transformation} in the main text.

Finally, to arrive at the polariton $T$ matrix and associated interaction constants, we need to consider how the matrix elements derived in Eq.~\eqref{eq:even partial wave scattering integral equation SM} relate to the single-particle eigenstates in the light-matter coupled system. For scattering between two polaritons $P_{m_1}$ and $P_{m_2}$, we define the associated polariton $T$ matrix as
\begin{equation}
    T^{(l)}_{P_{m_1}P_{m_2}}(k',k;E) \equiv \frac12\bra{P_{1},P_{2},k'l}\hat{T}(E)\ket{P_{1},P_{2},kl}.
\end{equation}
Using the correspondence between the eigenstates of the interaction and the bare polariton states, we find that the polariton $T$ matrix can be expressed as a linear superposition of the excitonic matrix elements
\begin{equation}
    T^{(l)}_{P_{m_1}P_{m_2}}(k',k;E) = \sum^{6}_{i,j=1} \bra{P_{m_1},P_{m_2},k'l}\ket{i,k'l}\bra{j,kl} \ket{P_{m_1},P_{m_2},kl} T^{(l)}_{ij}(k',k;E)
    \label{eq:polariton T-matrix SM}
\end{equation}
In order for the collision process to satisfy energy conservation, we have $k'=k$ at the end of the calculation, while the collision energy takes the form $E = E^{(m_1)}_{k} + E^{(m_2)}_{k}$.

The interaction constants are obtained by considering the scattering at $k=0$, in which case only $s$-wave scattering is nonzero. Therefore, for a pair of polaritons $P_{m_1}$ and $P_{m_2}$, we define
\begin{align} 
    g_{P_{m_1}P_{m_2}} \equiv \text{Re} \left[T^{(l=0)}_{P_{m_1}P_{m_2}}(0,0;E^{(m_1)}_{0} + E^{(m_2)}_{0}) \right].
\end{align}

\subsection{Scattering in the absence of light-matter coupling}
Finally, we discuss exciton scattering in the absence of light-matter coupling, as shown in Fig.~\ref{fig:Tx}(b) of the main text. Now the identity operator $\mathbb{1}$ and the non-interacting Hamiltonian $\hat{H}_{0}$ in Eq.~\eqref{eq:Green's operator SM} correspond to the case without the cavity photon mode. In this case, the non-interacting Hamiltonian can be diagonalized as
\begin{equation}
    \hat{H}_{0} = \sum_{\k}\left[ \sum^{2}_{\eta=1}E^{-}_{\k,\eta}\hat{L}^{\dagger}_{\k,\eta}\hat{L}_{\k,\eta} + \sum^{2}_{\eta=1}E^{+}_{\k,\eta}\hat{U}^{\dagger}_{\k,\eta}\hat{U}_{\k,\eta} \right],
    \label{eq:Hamiltonian without photon SM}
\end{equation}
where the operators $\hat{L}_{\k,\eta}$ and $\hat{U}_{\k,\eta}$, respectively, correspond to the lower and upper hybrid exciton modes that arise from DX$_{\eta}$-IX$_{\eta}$ hybridization for each $\eta$. The hybrid exciton dispersions are given by
\begin{equation}
    E^{\pm}_{\k,\eta} = \frac{1}{2}\left( \epsilon^{\text{IX}}_{\k,\eta} + \epsilon^{\text{DX}}_{\k} \pm \sqrt{(\epsilon^{\text{IX}}_{\k,\eta}-\epsilon^{\text{DX}}_{\k})^{2} + t^{2}} \right).
    \label{eq:hybrid exciton dispersions 2 SM}
\end{equation}
The diagonalization in Eq.~\eqref{eq:Hamiltonian without photon SM} can be achieved via the linear transformation
\begin{equation}
    \begin{pmatrix}
    \hat{L}_{\k,\eta} \\[0.2cm] \hat{U}_{\k,\eta}
    \end{pmatrix}
    = 
    \begin{pmatrix}
    u_{\k,\eta} & v_{\k,\eta} \\[0.2cm]
    -v_{\k,\eta} & u_{\k,\eta} \\
    \end{pmatrix}
    \begin{pmatrix}
    \hat{x}_{\k,\eta} \\[0.2cm] \hat{y}_{\k,\eta}
    \end{pmatrix},
\end{equation}
where the transformation coefficients take the analytic forms
\begin{equation}
    u^{2}_{\k,\eta} = \frac{1}{2} \left( 1 + \frac{\epsilon^{\text{IX}}_{\k,\eta}-\epsilon^{\text{DX}}_{\k}}{E^{+}_{\k,\eta}-E^{-}_{\k,\eta}} \right), \quad 
    v^{2}_{\k,\eta} = \frac{1}{2} \left( 1 - \frac{\epsilon^{\text{IX}}_{\k,\eta}-\epsilon^{\text{DX}}_{\k}}{E^{+}_{\k,\eta}-E^{-}_{\k,\eta}} \right),
\end{equation}
and satisfy $u^{2}_{\k,\eta}+v^{2}_{\k,\eta} = 1$.

We now define the matrix elements of the hybrid exciton propagator as $G^{X}_{in}(\q,E) = \bra{i,\q}\hat{G}(E)\ket{n,\q}$, using the two-particle states in Eq.~\eqref{eq:two-particle states SM}, the Green's operator in Eq.~\eqref{eq:Green's operator SM}, and the non-interacting Hamiltonian in Eq.~\eqref{eq:Hamiltonian without photon SM}. In the absence of light-matter coupling, hybrid exciton scattering decouples into two independent sectors ($\eta=1,2$), and we have the hybrid exciton propagators
\begin{equation}
    \mathbf{G}^{X}_\text{odd} = 
    \begin{pmatrix}
    G^{X}_{11} & G^{X}_{13} & G^{X}_{15} \\[0.2cm]
    G^{X}_{31} & G^{X}_{33} & G^{X}_{35} \\[0.2cm]
    G^{X}_{51} & G^{X}_{53} & G^{X}_{55}
    \end{pmatrix}, \quad 
    \mathbf{G}^{X}_\text{even} = 
    \begin{pmatrix}
    G^{X}_{22} & G^{X}_{24} & G^{X}_{26} \\[0.2cm]
    G^{X}_{42} & G^{X}_{44} & G^{X}_{46} \\[0.2cm]
    G^{X}_{62} & G^{X}_{64} & G^{X}_{66}
    \end{pmatrix}, 
\end{equation}
where
\begin{align}
    G^{X}_{11(22)}(\q,E) &= \frac{u^{4}_{\q,\eta}}{E-2E^{-}_{\q,\eta}+i0} + \frac{2u^{2}_{\q,\eta}v^{2}_{\q,\eta}}{E-E^{-}_{\q,\eta}-E^{+}_{\q,\eta}+i0} + \frac{v^{4}_{\q,\eta}}{E-2E^{+}_{\q,\eta}+i0}, \notag \\[0.2cm] 
    G^{X}_{13(24)}(\q,E) = G^{X}_{31(42)}(\q,E) &= \frac{u^{2}_{\q,\eta}v^{2}_{\q,\eta}}{E-2E^{-}_{\q,\eta}+i0} - \frac{2u^{2}_{\q,\eta}v^{2}_{\q,\eta}}{E-E^{-}_{\q,\eta}-E^{+}_{\q,\eta}+i0} + \frac{u^{2}_{\q,\eta}v^{2}_{\q,\eta}}{E-2E^{+}_{\q,\eta}+i0}, \notag \\[0.2cm] 
    G^{X}_{15(26)}(\q,E) = G^{X}_{51(62)}(\q,E) &= -\frac{\sqrt{2}u^{3}_{\q,\eta}v_{\q,\eta}}{E-2E^{-}_{\q,\eta}+i0} + \frac{\sqrt{2}u_{\q,\eta}v_{\q,\eta}(u^{2}_{\q,\eta}-v^{2}_{\q,\eta})}{E-E^{-}_{\q,\eta}-E^{+}_{\q,\eta}+i0} + \frac{\sqrt{2}u_{\q,\eta}v^{3}_{\q,\eta}}{E-2E^{+}_{\q,\eta}+i0}, \notag \\[0.2cm] 
    G^{X}_{33(44)}(\q,E) &= \frac{v^{4}_{\q,\eta}}{E-2E^{-}_{\q,\eta}+i0} + \frac{2u^{2}_{\q,\eta}v^{2}_{\q,\eta}}{E-E^{-}_{\q,\eta}-E^{+}_{\q,\eta}+i0} + \frac{u^{4}_{\q,\eta}}{E-2E^{+}_{\q,\eta}+i0}, \notag \\[0.2cm] 
    G^{X}_{35(46)}(\q,E) = G^{X}_{53(64)}(\q,E) &= -\frac{\sqrt{2}u_{\q,\eta}v^{3}_{\q,\eta}}{E-2E^{-}_{\q,\eta}+i0} - \frac{\sqrt{2}u_{\q,\eta}v_{\q,\eta}(u^{2}_{\q,\eta}-v^{2}_{\q,\eta})}{E-E^{-}_{\q,\eta}-E^{+}_{\q,\eta}+i0} + \frac{\sqrt{2}u^{3}_{\q,\eta}v_{\q,\eta}}{E-2E^{+}_{\q,\eta}+i0}, \notag \\[0.2cm] 
    G^{X}_{55(66)}(\q,E) &= \frac{2u^{2}_{\q,\eta}v^{2}_{\q,\eta}}{E-2E^{-}_{\q,\eta}+i0} + \frac{(u^{2}_{\q,\eta}-v^{2}_{\q,\eta})^{2}}{E-E^{-}_{\q,\eta}-E^{+}_{\q,\eta}+i0} + \frac{2u^{2}_{\q,\eta}v^{2}_{\q,\eta}}{E-2E^{+}_{\q,\eta}+i0},
    \label{eq:hybrid exciton propagator SM}
\end{align}
and $G^{X}_{in}=0$ otherwise. We use $\eta=1$ for the odd sector and $\eta=2$ for the even sector. Correspondingly, the $s$-wave scattering equation in Eq.~\eqref{eq:even partial wave scattering integral equation SM} decouples into the odd and even sectors as
\begin{equation}
    T^{(s)}_{ij}(k',k;E) = V^{(s)}_{ij}(k',k)\delta_{ij} + \sum_{n=\{1,3,5\}}\int^{\infty}_{0}\frac{q\,dq}{2\pi} V^{(s)}_{ii}(k',q)G^{X}_{in}(q,E)T^{(s)}_{nj}(q,k;E), \quad \text{$i,j \in \{1,3,5\}$},
\end{equation}
and 
\begin{equation}
    T^{(s)}_{ij}(k',k;E) = V^{(s)}_{ij}(k',k)\delta_{ij} + \sum_{n=\{2,4,6\}}\int^{\infty}_{0}\frac{q\,dq}{2\pi} V^{(s)}_{ii}(k',q)G^{X}_{in}(q,E)T^{(s)}_{nj}(q,k;E), \quad \text{$i,j \in \{2,4,6\}$}.
\end{equation}
The polariton $T$ matrix within the off-shell approximation can be expressed as a linear superposition of the off-shell excitonic matrix elements
\begin{align}
    T^{(s)}_{P_{m_1}P_{m_2}}(k',k;E) &= \sum_{i,j=\{1,3,5\}} \bra{P_{m_1},P_{m_2},k's}\ket{i,k's}\bra{j,ks} \ket{P_{m_1},P_{m_2},ks} T^{(s)}_{ij}(k',k;E) \notag \\[0.2cm] 
    &\quad + \sum_{i,j=\{2,4,6\}} \bra{P_{m_1},P_{m_2},k's}\ket{i,k's}\bra{j,ks} \ket{P_{m_1},P_{m_2},ks} T^{(s)}_{ij}(k',k;E),
\end{align}
where we take the limit $k' \to k$ while setting the collision energy by the corresponding polariton energies $E = E^{(m_1)}_{k} + E^{(m_2)}_{k}$.

\section{Stark-shift dependence of polariton interaction constants and Hopfield coefficients}

\begin{figure}[h]
    \centering
    \begin{minipage}{0.98\textwidth}
    \centering
    \includegraphics[width=\textwidth]{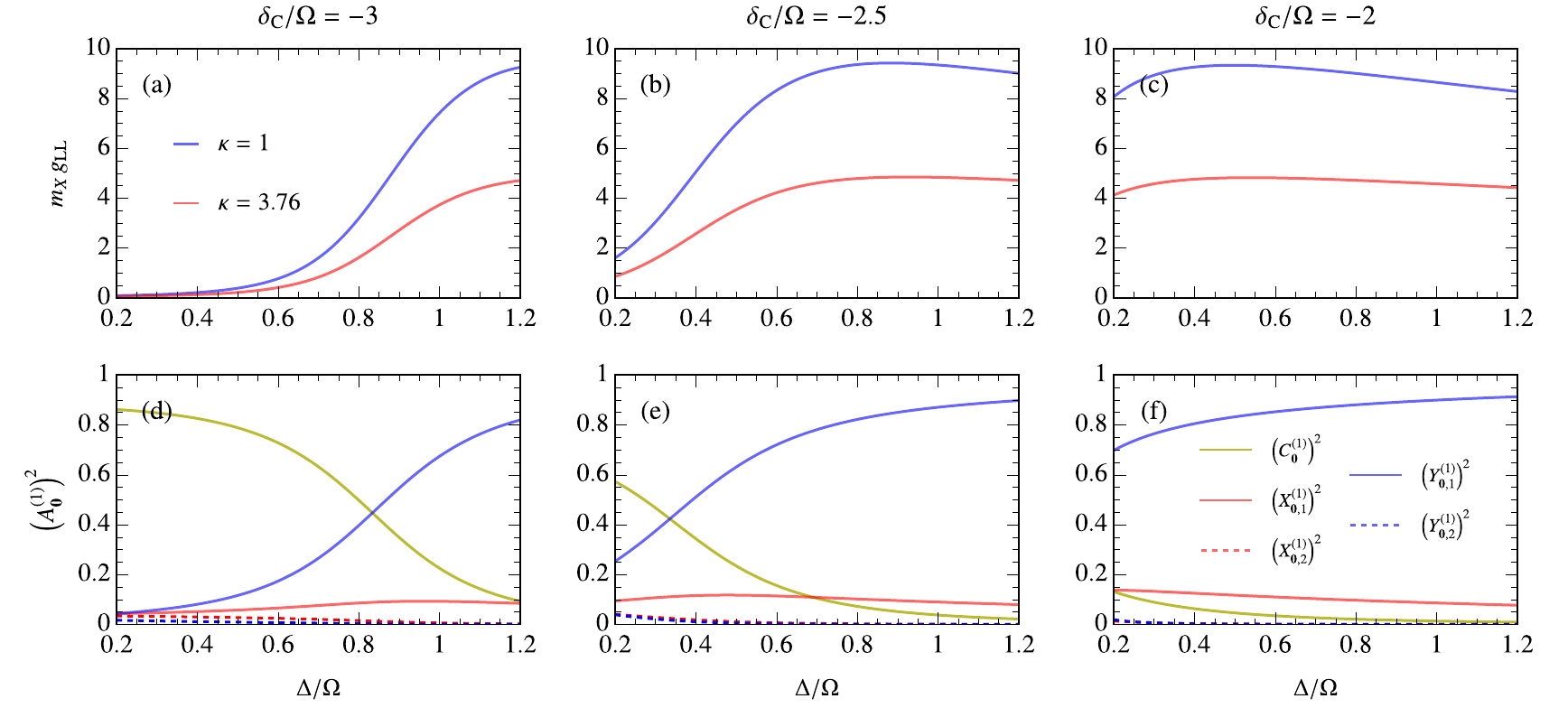}
    \end{minipage}
    \caption{(a,b,c) LP-LP interaction constant at zero momentum as a function of Stark shift. The blue and red lines, respectively, correspond to the case for vacuum ($\kappa=1$) and hBN ($\kappa=3.76$). (d,e,f) Corresponding Hopfield coefficients at zero momentum. The yellow, solid (dashed) red, and solid (dashed) blue lines, respectively, correspond to the photon, DX$_{1}$ (DX$_{2}$), and IX$_{1}$ (IX$_{2}$) fractions. Each column corresponds to a different value of the photon detuning, $\delta_\text{C}/\Omega = -3,-2.5,-2$ from left to right. In all cases, we use $\rho_{0}/a_{0} = 40$, $d/a_{0} = 6$, $\Omega/\varepsilon_\text{X} = 0.085$, $t/\varepsilon_\text{X} = 0.17$, and $\delta_\text{IX}/\varepsilon_\text{X} = -0.17$, where the reference scales $\rho_{0}$, $a_{0}$, and $\varepsilon_\text{X}$ correspond to $\kappa=1$.}
    \label{fig:TLL}
\end{figure}

In this section, we investigate how the lower polariton (LP) interaction constant $g_\text{LL}$ evolves as a function of Stark shift $\Delta$, which splits the two IX modes. In the following, we refer to the P$_{1}$ mode as the lower polariton (LP). This is of particular interest since the IX detuning can be controlled experimentally through an applied out-of-plane electric field~\cite{Cristofolini2012,Togan2018,Leisgang2020,Lorchat2021}, allowing one to tune the IX content of the LPs and thus their dipolar nature. Figures~\ref{fig:TLL}(a,b,c) show the LP interaction constant as a function of Stark shift for different photon detunings $\delta_\text{C}/\Omega=-3,-2.5,-2$. In each panel, we plot results for two values of the dielectric constant, corresponding to vacuum ($\kappa=1$) and hBN encapsulation ($\kappa=3.76$), shown in blue and red, respectively. As such, the two curves for $\kappa=1$ and $\kappa=3.76$ correspond to horizontal cuts of the density plots presented in Figs.~\ref{fig:gLL}(b,c) of the main text. Panels (d,e,f) show the corresponding photon, DX$_1$, DX$_2$, IX$_1$, and IX$_2$ fractions of the LP mode.

In general, we see that the LP interactions are largest when the exciton fraction is dominated by the IX component. This behavior agrees with the experimental observations of Ref.~\cite{Togan2018} in a microcavity hosting coupled InGaAs quantum wells, which reported an enhancement of the LP interactions as the indirect exciton fraction increased. However, we also observe in Figs.~\ref{fig:TLL}(a,b,c) that the LP interaction constant exhibits a peak at finite values of $\Delta$ and decreases for further positive detunings. This effect is due to the dependence of the interactions on the collision energy. Indeed, for large and positive $\Delta$, the LP approaches the lower hybrid exciton (hX$^{-}_{1}$) energy $E^{-}_{\mathbf{0},1}$ from below, and the interactions must vanish logarithmically when $E_\mathbf{0}^{(1)} \rightarrow E^{-}_{\mathbf{0},1}$. Overall, we find that the strongest LP interactions occur when the exciton content of the polariton is dominated by the indirect exciton (i.e., the interactions are enhanced when the polariton is more dipolar), while still maintaining a non-negligible photon fraction (i.e., the interactions are enhanced by the coupling to light).

\section{Polariton interaction constant without DX-IX interactions}

\begin{figure}[h]
    \centering
    \begin{minipage}{0.49\textwidth}
    \centering
    \includegraphics[width=\textwidth]{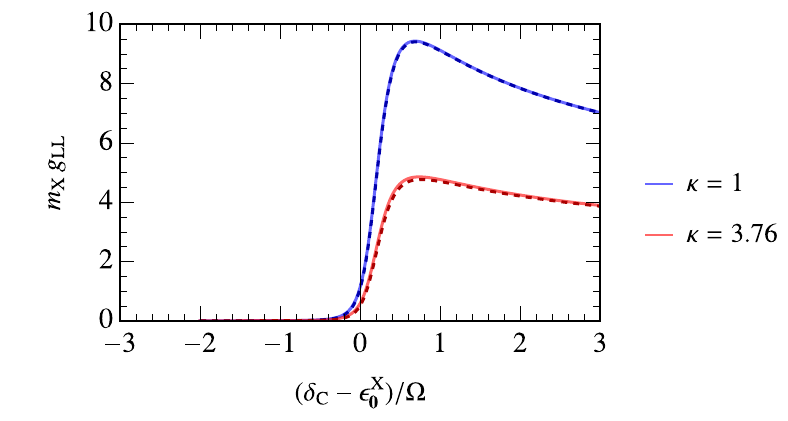}
    \end{minipage}
    \caption{LP-LP interaction constant at zero momentum as a function of photon detuning, where $\epsilon^\text{X}_\mathbf{0}=E^{-}_{\mathbf{0},1}$. The blue and red solid (dashed) lines correspond to the cases with (without) the DX-IX interactions for vacuum ($\kappa=1$) and hBN ($\kappa = 3.76$), respectively. In all cases, we use $\rho_{0}/a_{0} = 40$, $d/a_{0} = 6$, $\Omega/\varepsilon_\text{X} = 0.085$, $t/\varepsilon_\text{X} = 0.17$, $\delta_\text{IX}/\varepsilon_\text{X} = -0.17$, and $\Delta/\Omega = 1$, where the reference scales $\rho_{0}$, $a_{0}$, and $\varepsilon_\text{X}$ correspond to $\kappa=1$.}
    \label{fig:gLL without DX-IX SM}
\end{figure}

The dashed lines in Fig.~\ref{fig:gLL without DX-IX SM} show the LP interaction constant at zero momentum as a function of photon detuning in the absence of the DX-IX interactions, while the solid lines show the exact calculations including the DX-IX interactions, corresponding to Fig.~\ref{fig:gLL}(a) in the main text. Overall, we observe that the LP interaction constant is only marginally affected by the DX-IX interactions. This is mainly because the exciton fractions $2\left( X^{(1)}_{\mathbf{0},1} \right)^{2}\left( Y^{(1)}_{\mathbf{0},1} \right)^{2}$ and $ 2\left( X^{(1)}_{\mathbf{0},2} \right)^{2}\left( Y^{(1)}_{\mathbf{0},2} \right)^{2}$ applied to the DX-IX $T$ matrices in Eq.~\eqref{eq:polariton T-matrix SM} are generally smaller than those for the DX-DX and IX-IX interactions.


\end{document}